\documentclass[manuscript]{aastex}
\usepackage{rotating}
\usepackage{epsfig}
\newcommand{\Lsun}{\mbox{L$_\odot$}}
\newcommand{\Msun}{\mbox{M$_\odot$}}

\newcommand{\etal}{{ et~al.~ \/}}
\shortauthors{Zapata et al.}
\shorttitle{In Search of Disks Around Massive Stars}
\slugcomment{\it Submitted to The Astronomical Journal}
\begin{document}

\title{In Search of Circumstellar Disks Around Young Massive Stars}
\author{Luis A. Zapata\altaffilmark{1,2}, Luis F. Rodr\'\i guez\altaffilmark{1}, 
Paul T. P. Ho\altaffilmark{2,3}, Henrik Beuther\altaffilmark{2} 
and Qizhou Zhang\altaffilmark{2}}
\altaffiltext{1}{Centro de Radioastronom\'\i a y Astrof\'\i sica,
 UNAM, Apdo. Postal 3-72 (Xangari), 58089 Morelia, Michoac\'an, M\'exico}   
\altaffiltext{2}{Harvard-Smithsonian Center for Astrophysics, 60 Garden Street, 
Cambridge, MA 02138, USA}
\altaffiltext{3}{Academia Sinica Institute of Astronomy and Astrophysics,
Taipei, Taiwan.}

\email{lzapata@astrosmo.unam.mx}
 
\begin{abstract}
We present 7 mm, 1.3 cm and 3.6 cm continuum observations 
made with the Very Large Array
toward a sample of ten luminous IRAS sources that are believed to be 
regions of massive star formation.
We detect compact 7 mm emission in four of these objects: IRAS 18089-1732(1), 
IRAS 18182-1433, IRAS 18264-1152 and  IRAS 18308-0841 
and for the first time find that these IRAS sources are associated 
with double or triple radio sources separated by a few arcseconds.  
We discuss the characteristics of these sources based mostly
on their spectral indices and find that their nature is diverse.
Some features indicate that the 7 mm emission is dominated by dust 
from disks or envelopes. Toward other components the 7 mm emission 
appears to be dominated by free-free radiation, both from ionized outflows 
or from optically thick H II regions. Furthermore, there is evidence 
of synchrotron contamination in some of these sources.
Finally, we found that the sources associated 
with ionized outflows, or thermal jets are correlated with CH$_3$OH masers.
 The precise determination of the nature
of these objects requires additional multifrequency observations
at high angular resolution.   
The 3.6 cm continuum observations also revealed  
seven UCHII regions in the vicinity of the sources IRAS 18089-1732(1)
and two more in the source IRAS 18182-1433.
We show that the small photoionized nebulae of 
these UCHII regions are produced by early B-type stars.   

\end{abstract} 

\section{Introduction}
Massive stars (M$\geq$10~\Msun) are fundamental in determining the 
physical and chemical evolution of the galaxies. They produce a dominant fraction of 
the heavy elements, generate large amounts of UV radiation during their lives, 
and inject turbulent energy into the galactic ISM. However, the 
formation of massive stars is a poorly understood phenomenon. 
Schematically, there are two scenarios that have been proposed to explain the 
formation of massive stars. The first is the accretion scenario, 
(Garay \& Lizano 1998, Norberg \& Maeder 2000, and McKee \& Tan 2002)
 similar to that operating for low mass, where a dynamical collapse of molecular clumps,
results in the formation of disks and molecular outflows, which leads to the formation 
of a stellar core. In contrast to this, the coalescence scenario (Bonnell, Bate 
\& Zinnecker 1998; Stahler, Palla \& Ho  2000; and Bally \& Zinnecker 2005) 
 proposes that massive stars form by the merging of stars 
of low and intermediate masses. The possible existence of this coalescence scenario 
receives support from two facts: {\it i)} for a spherically symmetric geometry, 
the radiation pressure can reverse the infall process once the 
central star reaches about 10~\Msun~(Wolfire \& Cassinelli 1987; see, however, 
Yorke \& Sonnhalter 2002 for models that can produce stars with larger masses) 
and {\it ii)} most massive stars are born in the center 
of dense clusters of low and intermediate mass stars (Bonnell \etal 1998, 
and references therein). While the accretion scenario implies the existence 
of well formed disks and jets, this is not the case for in the coalescence 
scenario, where disks and jets are expected to be disrupted during 
the merging.

In recent years there has been a large effort to find 
observational evidence to discriminate between the two 
scenarios. Searches have been carried out to find
evidence for rotating disks using VLA ammonia observations. 
In two out of twelve objects studied, both B-type protostars, 
Zhang, Hunter \& Sridharan (1998) and Zhang et al. (2002) found
evidence for signatures of rotating disks. 
Pestalozzi et al. (2004) have interpreted VLBI observations of methanol masers
in NGC7538 IRS1 in terms of an edge-on Keplerian disk extending to a radius of 
$\sim$1000 AU and orbiting a 30 \Msun ~protostar. However, the methanol 
masers seem to be often associated with shocked molecular gas
(De Buizer 2003; Kurtz 2004). Gibb et al. (2004) have tried imaging 
dusty disks using subarcsecond BIMA millimeter 
observations (at 2.7 mm and 1.4 mm wavelengths) toward eight massive objects. 
In most of the objects they found that even at 2.7 mm the dominant
emission mechanism is free-free continuum from a thermal jet.
Moreover, if other OB-type stars have been formed previously in the region,
the object studied may be in the vicinity of bright free-free emission that will 
make it very difficult to search for the relatively weak mm emission from the disk. 
Finally, the shorter photoevaporation timescale of disks, the large distances 
($\geq$ Kpc), and the heavy extinctions (A$_V$ $\geq$ 100) have also been   
 inherently difficult problems in the search for disks around high-mass protostars. 

At present,  there is a list of early B-type  protostars that have 
been associated with possibles circumstellar disks, some of them are:  
the BN object (Jiang et al. 2005),  Cepheus A HW2 (Patel et al. 2005), IRAS 20126+4140 
(Cesaroni et al. 1997; 1999; Zhang, Hunter \& Sridharan 1998; Zhang et al. 2000),
IRAS 18089-1732 (Beuther et al. 2004b, 2005b), IRAS 18182-1433 (Beuther et al. 2006;
Zhang et al. 2006, both in preparation.),  G192.16-3.82 (Shepherd, Claussen, \& Kurtz 2001); 
W33A, AFGL 2591 \& NGC 7538 IRS9 (Van der Tak \& Menten 2005), AFGL~5142 (Zhang et al. 2002),
AFGL 490 (Schreyer et al. 2002 and 2005 in preparation).
 
We note that sources IRAS 18089-1732 and IRAS 18182-1433 are remarkable
examples of massive protostars that are associated with possible 
circumstellar disks that will be further discussed here.

In this paper, we present a centimeter and millimeter wavelength 
continuum study of ten IRAS sources that are thought to be regions 
of massive star formation. The purpose of this research 
is to identify the nature of these massive sources and 
to select a subset of good candidates for having associated 
a circumstellar disk.       
  
\section{The Sample}
       
The ten IRAS regions observed were selected from the sample of 
69 high mass protostellar objects (HMPOs) of Sridharan et al. (2002)
using the following criteria:
high luminosity ($\sim$2$\times$10$^4~L_\odot$) suggesting the
presence of OB-type protostars;
compact size in their mm emission 
($\leq$ 20${''}$$\times$20${''}$); proximity to the Sun (a few Kpc or less);
bright flux density at 1.2 mm ($\geq$200 mJy), and weak centimeter emission ($\leq$1 mJy).
The criterion of weak centimeter emission was included to search
for sources where the continuum emission at 7 mm and shorter wavelengths
from possible disks is less likely to be contaminated by free-free emission.
In Table 1 we show the main properties of the ten objects 
selected.

\section{Observations}
 
The observations were made with the Very Large Array of 
NRAO\footnote{The National Radio Astronomy Observatory is a 
facility of the National Science Foundation operated
under cooperative agreement by Associated Universities, Inc.} 
in the continuum mode at 7 mm during 2004 April 1, and at 1.3 cm 
and 3.6 cm on 2004 May 17. At these epochs, the VLA was in its C 
configuration. The absolute flux calibrator was 1331+305. 
In Tables 2, 3 and 4 we present a summary of the observations 
that will be discussed here, including the measured flux density of 
the phase calibrators. 

As mentioned above, during the first epoch, we made 7 mm continuum 
observations toward the ten selected massive young regions.  We 
detected continuum emission at or above a 3-$\sigma$ 
level of $\sim$0.5 mJy in four of the regions: IRAS 18089-1732(1), 
IRAS 18182-1433, IRAS 18264-1152, and IRAS 18308-0841. 
During the second epoch, we observed at 1.3 and 3.6 cm wavelengths 
toward these four regions.

The data were edited and calibrated in the standard manner
using the software package Astronomical Image Processing 
System (AIPS) of NRAO. 
Clean maps were obtained using the task IMAGR of AIPS. 
For the 7 mm continuum maps we used the ROBUST parameter set to 5,
that corresponds to natural weighting, to achieve maximum 
sensitivity in each continuum image. 
In the 1.3 and 3.6 cm continuum maps we used the ROBUST parameter set to 0,
for optimal compromise between sensitivity and angular resolution.
The contour and grey scale maps at 0.7, 1.3 and 3.6 cm of 
the four regions are shown in Figures 1, 2, 4, 5, 7, 8, 10, and 11.  
The resulting rms noises and synthesized beam parameters for the
 0.7, 1.3 and 3.6 cm continuum images are also given in Tables 2, 3 and 4. 

Additionally, in all 7 mm maps we have also applied a 
uv-tapering between 100 to 200 k$\lambda$ in order to increase
 the signal-to-noise ratio of the extended emission ($\geq$ 1${''}$) 
 in these regions.
  
\section{Results and Discussion.}

In what follows, we discuss each of the four sources detected at 7 mm separately.
Our discussion of the spectral indices is based on the assumption that the 
flux densities did not change between the 2004 April 1 and 2004 May 17 observations.

%%%%%%%%%%%%%%%%%% SOURCE IRAS 18089-1732

\subsection{IRAS 18089-1732(1)}

This source is previously known as IRAS 18089-1732, however we have added
a numbering (1) at the end of the name to distinguish between this
component and component IRAS 18089-1732(4)

This region contains a strong and extended ($\sim$ 11${''}$) 
1.2 mm continuum source, H$_2$O and Class II CH$_3$OH maser spots, 
and a compact ($\sim$ 1${''}$) 3.6 cm continuum source that is centered 
at the middle of the extended 1.2 mm continuum source (Beuther et al. 2002a, 2002c).
In recent millimeter and submillimeter observations made with the
Submillimter Array (SMA), Beuther et al. (2004b, 2005b) found a 1.2 mm continuum compact 
core elongated in the southeast-northwest direction, 
while in the 860 $\mu$m continuum observations they detected 
emission more compact, that is coincident with 
the 3.6 cm source and the H$_2$O maser (see Beuther 2004b, Figure 1). 
Furthermore, Beuther et al. (2004b, 2005b) found a molecular line forest 
associated with this source, and particularly the HCOOCH$_3$(20-19) 
line appears to show evidence for the signatures of  rotation. 
Finally, IRAM 30 m single-dish SiO J=2$\rightarrow$1 (Beuther, priv. comm.), and 
SMA SiO J=5$\rightarrow$4 (Beuther et al. 2004b) 
observations show evidence of outflowing molecular gas emanating from this region.     
 
The 7 mm maps (see Figures 1 and 2) show that this source is actually a double
source with components separated approximately by $\sim$ 2$''$ (7200 AU, at a 
distance of 3.6 Kpc) in the northeast and southwest direction. We refer to these 
components as IRAS 18089-1732(1)a and IRAS 18089-1732(1)b. 

In Figure 2 we have overlayed the 1.2 and 0.86 mm continuum maps 
from Beuther et al. (2004b) with our 7 mm continuum maps. 
Figures 2a and 2b show that the 1.2 and 0.87 mm continuum cores found by
Beuther et al. are associated with the compact 7 mm continuum source 
IRAS 18089-1732(1)a. We now discuss each subcomponent separately.

\subsubsection{IRAS 18089-1732(1)a}

The radio source IRAS 18089-1732(1)a is resolved in the 7 mm maps.
It has deconvolved dimensions of $2\rlap.{''}3$ $\pm$ 
$0\rlap.{''}4$ $\times$ $1\rlap.{''}1$  
$\pm$ $0\rlap.{''}4$ and a P.A. $= 28^\circ \pm 15^\circ$. 
In the 1.3 cm continuum maps we detect the counterpart of
IRAS 18089-1732(1)a (see Figure 1b). This source is also resolved
and has deconvolved dimensions of $1\rlap.{''}4$ $\pm$ $0\rlap.{''}2$ 
$\times$ $0\rlap.{''}9$  $\pm$ $0\rlap.{''}2$ and a 
P.A. $= 32^\circ \pm 13^\circ$. 
The position angle of the source is similar at 7 mm and
1.3 cm, as well as to the position angle of the monopolar SiO outflow
reported by Beuther et al. (2004a).

Also the 3.6 cm continuum map shows a source that is coincident 
with IRAS 18089-1732(1)a (see Figure 1a), however, with our resolution 
($\sim$ 4${''}$) we cannot resolve it. 
From figure 1 of Beuther et al. (2004a) and our Figures 1 and 2, 
we associate this 3.6 cm continuum source with the source found previously by 
Beuther et al. (2002b) at the same wavelength. 
Moreover, the flux densities for both epochs are quite similar, $\sim$1 mJy.  

Figure 3a shows the spectral energy distribution for IRAS 18089-1732(1)a from
the centimeter to submillimeter wavelengths. It shows a ``combined'',
two-regime spectrum in which the centimeter emission is dominated by a flat or slowly rising 
spectrum ($\alpha$=0.58$\pm$0.05) which can be interpreted as moderately optically thick 
free-free emission, while the millimeter and sub-millimeter emission is dominated 
by a component that rises rapidly with frequency. This component is likely to be 
associated with dust emission from a core or disk. 

Since the orientation of the position angle
of IRAS 18089-1732(1)a is in the south-north direction (in the 1.3 cm and 7 mm observations), 
and its spectral energy distribution at centimeter wavelengths suggests that 
this source may be associated with a thermal jet or stellar wind,
we interpret this source as a  
thermal jet that is possibly driving the north-south
SiO molecular outflow found by Beuther et al. (2004b). Moreover, as we will see in our sample, 
these free-free emission sources of relatively flat spectrum appear to be systematically
associated with CH$_3$OH maser spots (see Figure 1b and discussion below).   
      
\subsubsection{IRAS 18089-1732(1)b}
 
This source is reported here for the first time.
IRAS 18089-1732(1)b is resolved in the 7 mm maps.
It has deconvolved dimensions of $2\rlap.{''}7$ $\pm$ 
$0\rlap.{''}2$ $\times$ $1\rlap.{''}6$  
$\pm$ $0\rlap.{''}2$ and a P.A. $= 35^\circ \pm 46^\circ$. 

IRAS 18089-1732(1)b is only detected at 7 mm.
Figure 3b shows its spectral energy distribution from
the centimeter to submillimeter wavelengths, mostly upper limits. 
From our figures and the SED this source could be interpreted
as associated with dust emission. 
However, the spectrum of this source seems 
to flatten between 1.2 mm and 850 $\mu$m (see Figure 3
and Beuther et al. 2004b).
Therefore, we speculate that IRAS 18089-1732(1)b could be also an
optically thick H~II region that has its turnover frequency at
about 100 GHz.

%%%%%%%%%%%%%%%%%% SOURCE IRAS 18182-1433 %%%%%%%%%%%%%%%%%%

\subsection{IRAS 18182-1433}
  
This region contains a strong and extended ($\sim$ 13${''}$) 
1.2 mm continuum source and H$_2$O and Class II CH$_3$OH masers spots 
(Beuther et al. 2002a, 2002c). In recent SMA millimeter continuum and line 
observations (Beuther et al. 2005, in preparation), a compact core 
($\sim$ 4$''$) has been detected that is also associated with a molecular line forest. 
This millimeter continuum core is associated with the H$_2$O masers. 
Beuther et al. (2002b) also find an extended ($\sim$ 1$'$)  
CO(2-1) molecular outflow emanating from this region.
     
The 3.6, 1.3 and 0.7 cm maps (see Figures 4 and 5) show that IRAS 18182-1433
is a triple source with components separated by 2 to 10 arcseconds.
We refer to these components as IRAS 18182-1433a, IRAS 18182-1433b and 
IRAS 18182-1433c. 

\subsubsection{IRAS 18182-1433a}
 
IRAS 18182-1433a is detected in our 7 mm continuum map, but not in our 
3.6 cm and 1.3 cm maps above the 4$\sigma$ level of 55 and 70 $\mu$Jy, 
respectively. This source is not resolved in the 7 mm continuum map. However,
based on our synthesized beam, we set an upper limit of $\leq$ 2 ${''}$ 
for its angular size.   

In Figure 5 we can see that this source is associated with the
water masers and the 1.2 mm continuum core
found by Beuther et al. (2002a,c). However, there is a shift of 
about 0.3$''$ between the two peaks at 7 and 1.2 mm. One possibility, is that    
this shift could be due to the fact that dust emission contribution 
at 1.2 mm comes from both sources, IRAS 18182-1433a and IRAS 18182-1433b, 
while the source IRAS 18182-1433b was not detected in our 7 mm continuum map.

Figure 6a shows the spectral energy distribution for IRAS 18182-1433a from
the centimeter to submillimeter wavelengths. It shows a spectrum associated 
with a strong dust emission, possibly from the core and disk.

\subsubsection{IRAS 18182-1433b}

IRAS 18182-1433b is resolved in our 1.3 cm continuum maps.
It has deconvolved dimensions of $1\rlap.{''}7$ $\pm$ 
$0\rlap.{''}4$ $\times$ $1\rlap.{''}4$ $\pm$ $0\rlap.{''}6$ 
and a P.A. $= 87^\circ \pm 40^\circ$. 

IRAS 18182-1433b is not detected in our 7 mm map above 
a 4$\sigma$ level of 0.22 mJy. 
The CH$_3$OH Class II maser spot found by Beuther et al. 
(2002c) is associated with this source.
Figure 6b shows its spectral energy distribution from
the centimeter to submillimeter wavelengths.  
It shows a slowly rising  ($\alpha$=0.10$\pm$0.01) spectrum.
 We propose that is produced by free-free emission from  
a thermal jet or stellar wind. However, this flat spectral index 
is also consistent with an optically thin H II region.

\subsubsection{IRAS 18182-1433c}

This source is not resolved in our 1.3 cm maps. Again, based on 
our synthesized beam, we set an upper limit of $\leq$ 2${''}$ 
for its angular size. IRAS 18182-1433c is not detected in our 7 mm map above 
a 4$\sigma$ level of 0.22 mJy.
 
Figure 6c shows its spectral energy distribution from
the centimeter to submillimeter wavelengths. 
It shows a slightly negative spectrum ($\alpha$=-0.36$\pm$0.05), 
which is suggestive of non thermal radiation.
Similar negative spectral indices found in sources
associated with massive and low-mass star formation are believed to
be due to synchrotron contamination from strong shocks
(Serpens: Rodr\'\i guez et al. 1989, HH~80-81:
Mart\'\i, Rodr\'\i guez, \& Reipurth 1993, Cep A: Garay et al. 1996,
W3(OH): Wilner, Reid, \& Menten 1999, IRAS~16547-4247: Rodr\'\i guez et al. 2005
G192.16-3.82: Shepherd \& Kurtz 1999). 
A detailed discussion on the presence of thermal and non thermal 
components in this type of objects is given in Garay et al. (1996).

%%%%%%%%%%%%%%%%%% SOURCE IRAS 18264-1152 %%%%%%%%%%%%%%%%%%

\subsection{IRAS 18264-1152}

This region contains a strong and extended ($\sim$ 15${''}$) 
1.2 mm continuum source and H$_2$O and Class II CH$_3$OH maser spots 
(Beuther et al. 2002a, 2002c). IRAS 18264-1152 shows evidence of 
outflowing activity. Beuther et al. (2002b) found an extended ($\sim$ 1$'$) 
CO(2-1) molecular outflow emanating from here.
      
The 7 mm and 1.3 cm maps (see Figures 7 and 8) show that this source is a 
triple source, with components separated approximately by $\sim$2$''$. 
We refer to these components as IRAS 18264-1152a, IRAS 18264-1152b, and 
IRAS 18264-1152c.  We propose that they are possibly tracing the components 
of a triple stellar system.

\subsubsection{IRAS 18264-1152a}

Figure 9 shows the spectral energy distribution of this source from
the centimeter to millimeter wavelengths.
The power law index of 2.1$\pm$0.05 suggests an optically thick H~II region
or dust emission from the core and disk.

\subsubsection{IRAS 18264-1152b}

This source is resolved in our 7 mm continuum maps.
It has deconvolved dimensions of $3\rlap.{''}8$ $\pm$ 
$0\rlap.{''}5$ $\times$ $1\rlap.{''}4$ $\pm$ $0\rlap.{''}4$ 
and a P.A. $=143^\circ \pm 7^\circ$.

Figure 9 shows its spectral energy distribution from
the centimeter to millimeter wavelengths.
The slightly rising spectrum (spectral index of 0.27$\pm$0.06)
between 3.6 and 1.3 cm suggests a thermal jet or a partially
optically thick H~II region. The flux density at 7 mm is
much larger than the value extrapolated from the 3.6 and 1.3 cm
observations and suggests that dust emission may be dominant at
7 mm.

\subsubsection{IRAS 18264-1152c}

This source is resolved in our 7 mm continuum maps.
It has deconvolved dimensions of $2\rlap.{''}4$ $\pm$ 
$0\rlap.{''}5$ $\times$ $0\rlap.{''}9$ $\pm$ $0\rlap.{''}7$ 
and a P.A.$= 45^\circ \pm 13^\circ$.

Figure 9 shows its spectral energy distribution from
the centimeter to millimeter wavelengths.
At 3.6 cm, we have only an upper limit. The 
power law index of 2.1$\pm$0.05 between 1.3 cm and 7 mm suggests
an optically thick H~II region or dust emission from 
a core and disk.

%%%%%%%%%%%%%%%%%% SOURCE IRAS 18264-1152 %%%%%%%%%%%%%%%%%%%%%

\subsection{IRAS 18308-0841}

This region contains a strong and extended ($\sim$ 17${''}$) 
1.2 mm continuum source and one H$_2$O maser spot (Beuther et al. 2002a, 2002c). 
Beuther et al. (2002b) found strong $^{12}$CO(2-1) molecular emission. However
the presence of bipolar outflow structure was not clearly established.

The 7 mm and 1.3 cm maps (see Figures 10 and 11) show that this source 
is a triple source with components separated by 3$''$ to 10$''$. 
We refer to these components as IRAS 18308-0841a, IRAS 18308-0841b 
and IRAS 18308-0841c.

\subsubsection{IRAS 18308-0841a}

This source is resolved in our 7 mm continuum maps.
It has deconvolved dimensions of $2\rlap.{''}16$ $\pm$ 
$0\rlap.{''}04$ $\times$ $2\rlap.{''}0$  
$\pm$ $0\rlap.{''}05$ and a P.A. $= 92^\circ \pm 14^\circ$. 
This source is also resolved in our 1.3 cm continuum maps.
It has deconvolved dimensions of $1\rlap.{''}3$ $\pm$ 
$0\rlap.{''}9$ $\times$ $0\rlap.{''}7$  
$\pm$ $0\rlap.{''}2$ and a P.A. $= 100^\circ \pm 8^\circ$. 
Finally, the source is also resolved in our 3.6 cm continuum maps.
It has deconvolved dimensions of $2\rlap.{''}4$ $\pm$ 
$0\rlap.{''}1$ $\times$ $2\rlap.{''}20$  
$\pm$ $0\rlap.{''}05$ and a P.A. $= 60^\circ \pm 16^\circ$. 
We find that while the 7 mm and 1.3 cm deconvolved dimensions are
consistent, the deconvolved dimensions at 3.6 cm appear
to be significantly larger. This may be due to the presence of
extended emission that is resolved out or too faint to be detected at
7 mm and 1.3 cm.

Figure 12 shows its spectral energy distribution from
the centimeter to millimeter wavelengths.
The negative spectral index of -0.42$\pm$0.07 could be suggest the presence 
of synchrotron emission. However, this spectral index could 
also be due to missing flux density in the higher angular resolution and
higher frequency observations. Matching beam observations of the source at
several frequencies are needed to discuss the spectral indices
in a very reliable way.

\subsubsection{IRAS 18308-0841b}

This source is resolved in our 3.6 cm continuum maps.
It has deconvolved dimensions of $3\rlap.{''}1$ $\pm$ 
$0\rlap.{''}3$ $\times$ $2\rlap.{''}5$  
$\pm$ $0\rlap.{''}2$ and a P.A. $= 75^\circ \pm 12^\circ$. 

Figure 12 shows its spectral energy distribution from
the centimeter to millimeter wavelengths.
Again, the negative spectral (-0.26$\pm$0.01) index could be suggest
the presence of synchrotron emission. However, this spectral index could 
also be due to missing flux density in the higher angular resolution and
higher frequency observations. Matching beam observations of the source at
several frequencies are also needed to discuss the spectral indices
in a very reliable way.

\subsubsection{IRAS 18308-0841c}

Figure 12 shows its spectral energy distribution from
the centimeter to millimeter wavelengths.
It was detected only at 7 mm, suggesting that 
dust is dominant at this wavelength.

\subsection{Clusters of UCHII Regions in the Vicinity of the 
sources IRAS 18089-1732(1) and IRAS 18182-1433}

We report for the first time the detection of seven ultracompact HII regions
in the vicinity of the source IRAS 18089-1732(1) and 
two more ultracompact HII regions in the source IRAS 18182-1433 at 3.6 cm 
(see Figure 13 and 14).
Their physical parameters are shown in Table 5. Most of them show  
a cometary or unresolved spherical morphology. Following Van der Tak
and Menten (2005), and assuming that the free-free emission is optically thin, 
that the ionized gas has a temperature of 10$^4$ K, and that these sources have 
distances of 3.6 and 4.5 kpc, we find that the flux of Lyman continuum photons 
for each UCHII, can be provided by an early B-type star 
(see Panagia 1973).

We searched for more UCHII regions in the vicinity of the others two IRAS sources that were
observed at 3.6 cm, however, we did not detect other cases at a 4$\sigma$ level of their
respective rms noises. Therefore, in comparison with these regions, 
IRAS 18089-1732(1) and IRAS 18182-1433 can be considered rich regions 
of high mass star formation.     
 
\subsection{General Discussion of the Sources}
  
With the present angular resolution and sensitivities of our images, it is 
interesting to note that the four sources detected at 7 mm (IRAS 18089-1732, 
IRAS 18182-1433, IRAS 18264-1152, and IRAS 18308-0841) 
are double or triple, with no case of a single source.  
In the four regions with detected 7 mm emission, we find a total of 11 subcomponents.
The spectral index of these subcomponents varies and suggests different
interpretations for the nature of the sources.
In Table 6 we give a tentative interpretation for the nature of
the sources, based on our discussion. 

We note that the sources IRAS 18089-1732(1)a, IRAS 18182-1433b, IRAS 18264-1152b  
 hat exhibit slightly rising spectral indices (that are frequently associated with winds 
or thermal jets), show a correlation with methanol masers spots. Furthermore, 
those maser spots are found in the same orientation of the molecular 
outflows (See figure 1b, 5 and 8).
Since the methanol masers already have been 
suggested to be associated with shocked molecular gas in the outflows produced
by the star (De Buizer 2003; Kurtz 2004), our data show this correlation, and supports 
the notion that the methanol masers are associated with ionized or shocked gas 
rather than disks (or dust emission).     

It is puzzling that we only detected 4 of the 10 sources observed, since
they all (with the exception of IRAS 18553+0414 and IRAS 18089-1732(4)) seem 
to have similar luminosities, and distances. However, these four detected continuum
 sources (IRAS 18089-1732, IRAS 18182-1433, IRAS 18264-1152 and IRAS 18308-0841) 
are embedded in the most massive cores of our initial subset (see Table 1).

Of the 11 subcomponents detected, four (IRAS 18089-1732(1)b, 
IRAS 18182-1433a, IRAS 18264-1152b, IRAS 18308-0841c) 
have 7 mm emission that could be coming from the dust. 

\section{Conclusions}

Our main conclusions are summarized below.

1) We mapped ten regions of massive star formation at 7 mm
with the VLA, detecting compact emission in four of them.
These four sources were further observed with
the VLA at 1.3 and 3.6 cm. 

2) Multiplicity is a common factor in the emission detected:
three of the four sources detected are resolved into three subcomponents,
while the fourth source is a double source.

3) Four of the sources detected show 7 mm emission above the value
extrapolated from the observations at longer wavelengths assuming jet emission. 
These sources are IRAS 18089-1732(1)b, IRAS 18182-1433a, IRAS 18264-1152b,
IRAS 18308-0841c. They are considered good candidates of massive young stars
with dust emission at 7 mm that could come from a disk or envelope.
However, further observations at higher and lower frequencies are required 
to establish if the detected 7 mm emission is actually coming from 
dust or from optically thick ionized gas (free-free emission).

4) Two of the sources show spectral indices consistent with
ionized outflows: IRAS 18089-1732(1)a and IRAS 18182-1433b.
IRAS 18264-1152b show, in addition to the 7 mm excess, a
spectral index consistent with an ionized outflow between
3.6 and 1.3 cm. This last source may then be a combination of thermal jet
and disk. 
 
5) Two of the sources show spectral indices consistent
with optically thick HII regions: IRAS 18264-1152a and c.

6) Three of the sources show negative spectral
indices that suggest the presence of synchrotron contamination:
IRAS 18182-1433c, and IRAS 18308-0841a and b.
However, these spectral indices could also be due to missing
flux density in the higher angular resolution, higher frequency
observations. If confirmed with multifrequency, matching beam observations,
this synchrotron emission could be an indicator of strong shocks
produced by the powerful outflows known to exist in these
regions.

8) Finally, we also find that the methanol masers in these regions are systematically
associated with radio sources that show slightly rising spectral indices:
IRAS 18089-1732(1)a, IRAS 18182-1433b, and IRAS 18264-1152b. This indicates 
that the methanol maser emission is associated with outflows rather than with
disks. The interpretation is that these masers are associated with the presence 
of compact free-free emission.

Our findings are consinstent with the accretion scenario, where 
a dynamical collapse of molecular clumps, results in the formation of disks 
and molecular outflows, which leads to the formation of a massive stellar core.
However, in each IRAS source we also detected multiple radio sources separate 
a few arcseconds, suggesting the formation of groups of massive stars.

\acknowledgments
 L.A.Z. acknowledges the Smithsonian Astrophysical Observatory 
for a predoctoral fellowship. L.F.R. acknowledges the support
of DGAPA, UNAM, and of CONACyT (M\'exico).
H.B. acknowledges financial support by the Emmy-Noether-Programm of the
Deutsche Forschungsgemeinschaft (DFG, grant BE2578/1).
This research has made use of the SIMBAD database, 
operated at CDS, Strasbourg, France.

%%%%%%%%%%%%%%%%%% SOURCE IRAS 18089-1732

\begin{figure}
\vspace{-1.2cm}
\begin{center}
\scalebox{.50}{\includegraphics{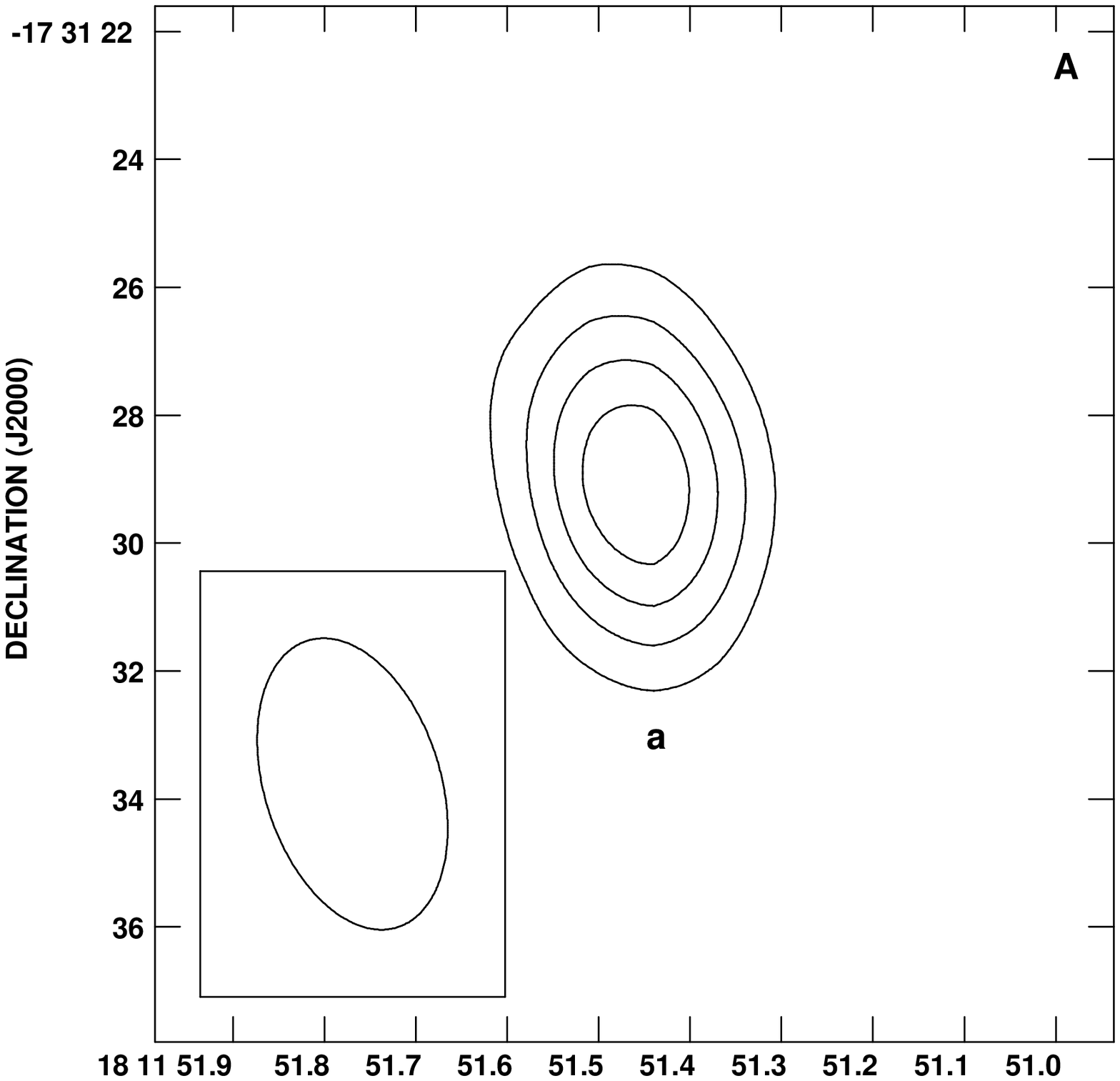}}\vspace{-1.8cm}\\
\scalebox{.50}{\includegraphics{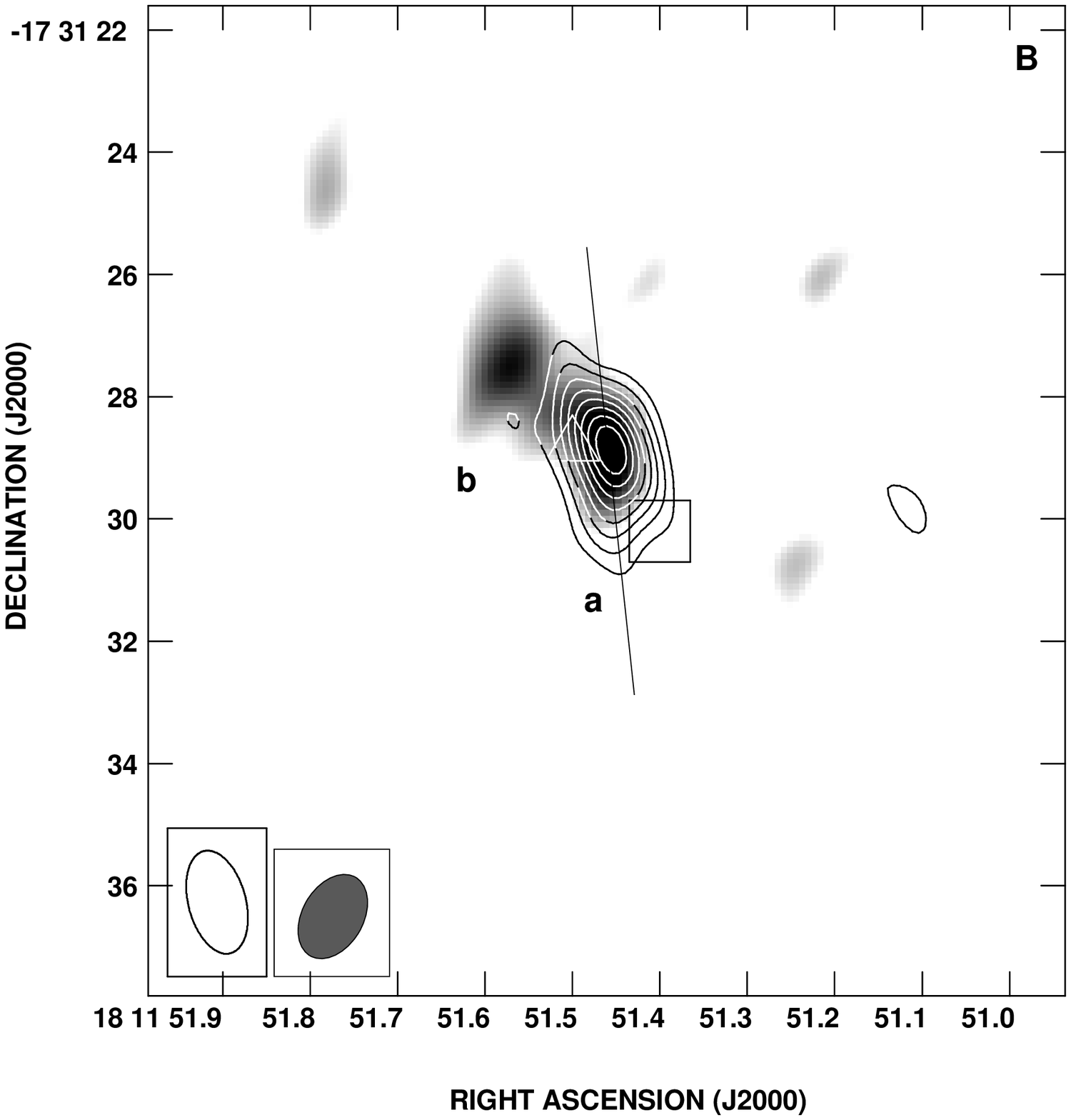}}\vspace{-1.5cm}\\
\end{center}
\vspace{0.5cm}
\caption{{\scriptsize VLA continuum images of the source IRAS 18089-1732(1).
 {\bf A}). The 3.6 cm continuum emission. The contours are -4, 4, 6, 8, and 10 times 40 $\mu$Jy 
beam$^{-1}$, 
 the rms noise of the image. {\bf B).} The 1.3 cm continuum emission (contours) and 7 mm continuum 
emission (grey scale).
 The contours are -4, 4, 6, 8, 10, 12, 14, 16 and 18 times 70 $\mu$Jy beam$^{-1}$, the rms noise of the image. 
 The half power contour of the synthesized beam is shown in the bottom left corner of each imagen.
 The 7 mm continuum map has a (u,v) tapering of 200 k$\lambda$.
 The square and triangle indicate CH$_{3}$OH and H$_{2}$O masers positions (Walsh et al. 1998 and Beuther et al. 2002),
 respectively. The line indicates the direction of the SiO molecular outflow found by Beuther et al. 
(2004b).}}
\label{fig1.0}
\end{figure}

\newpage 
\begin{figure}
\vspace{-0.5cm}
\begin{center}
\scalebox{.50}{\includegraphics{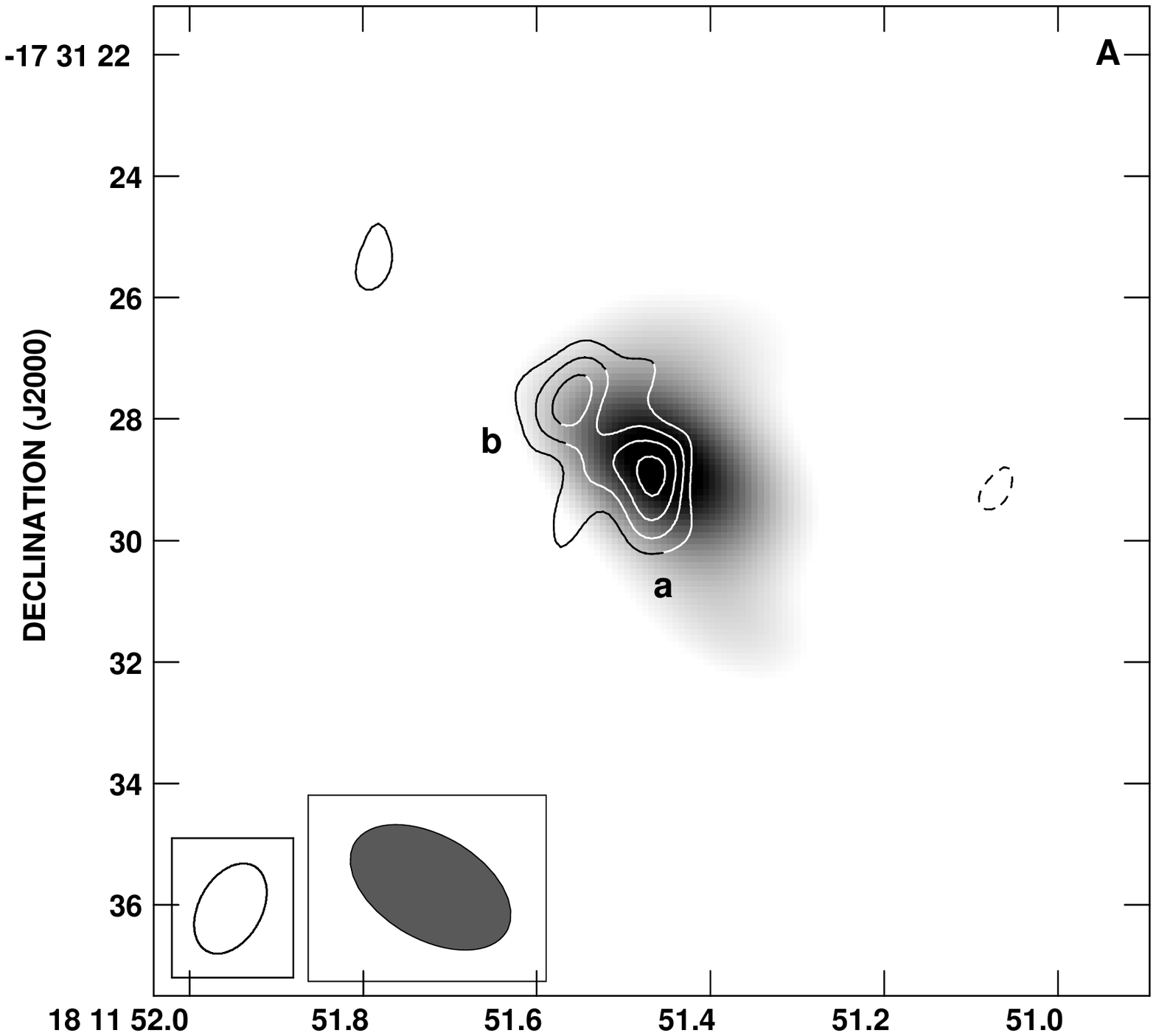}}\vspace{-0.5cm}\\
\scalebox{.50}{\includegraphics{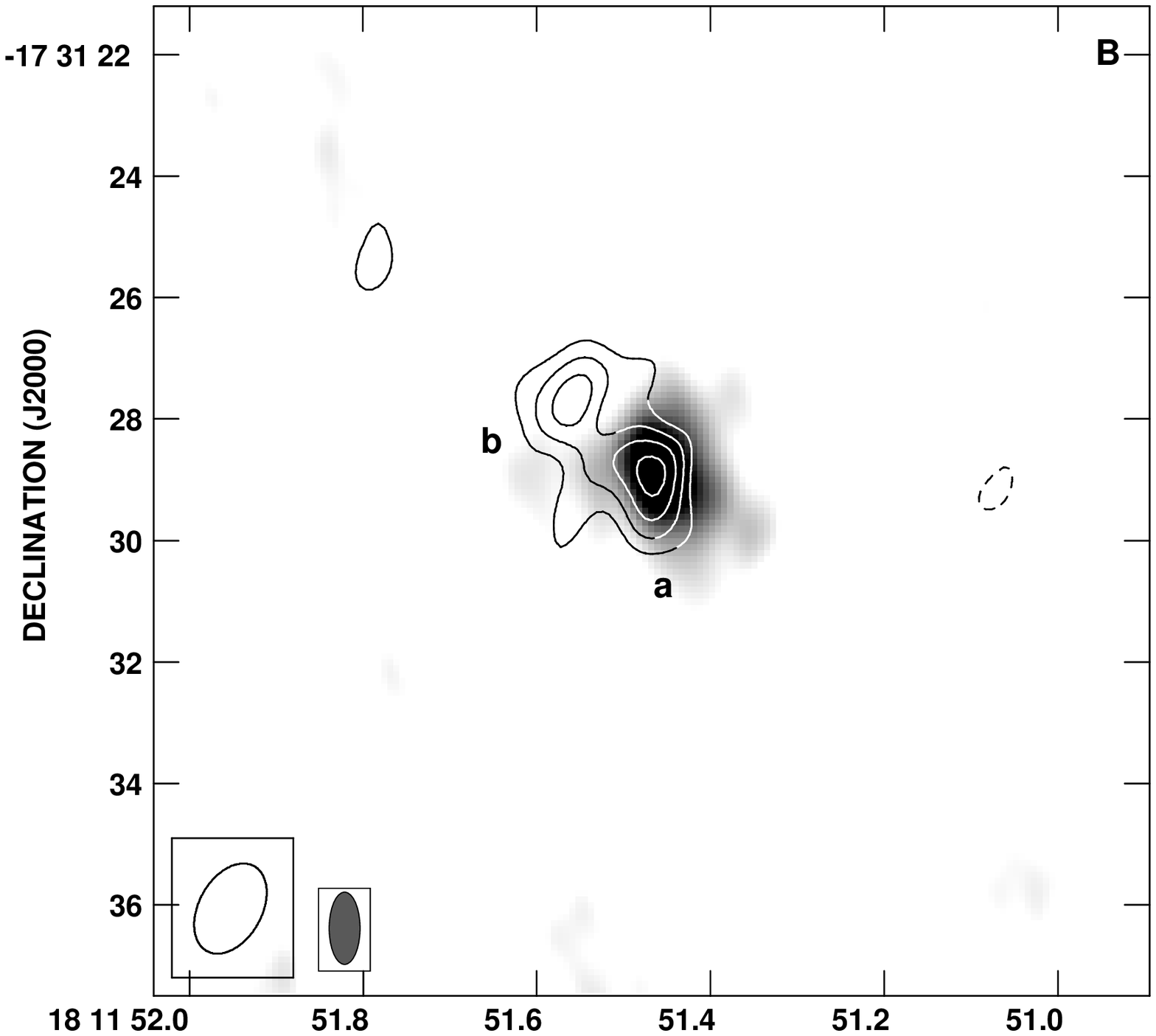}}\vspace{-0.9cm}\\
\end{center}
\vspace{0.5cm}
\caption{{\scriptsize VLA and SMA continuum images of the source IRAS 18089-1732(1).
The contours shows the 7 mm continuum emission and the grey scale shows the SMA continuum 
emission at 1.2 mm ({\bf A}) and 860 $\mu$m ({\bf B}) (Beuther et al. 2004b,2005). 
The half power contour of the synthesized beam is 
shown in the bottom left corner of each image. 
In the {\bf A} and {\bf B} images the contours are -4, 4, 5, 6, 7 and 8 times 0.19 mJy 
beam$^{-1}$, the rms noise of the image. The 7 mm continuum map has a (u,v) tapering 
of 150 k$\lambda$.}}
\label{fig1.1}
\end{figure}

\newpage 
\begin{figure}
\begin{center}
\scalebox{.80}{\includegraphics{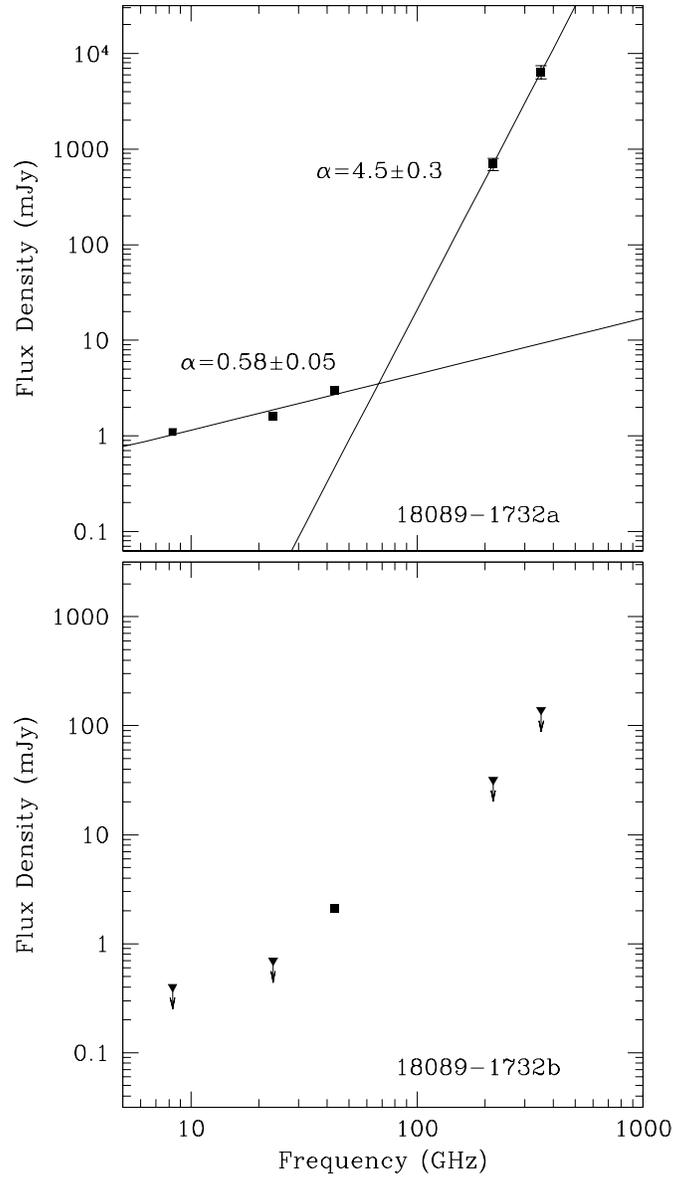}} 
\caption{{\scriptsize Spectral energy distribution (SED) for each detected radio continuum 
source in IRAS 18089-1732(1) combining the 0.7, 1.3 and 3.6 cm VLA continuum data and 
the SMA continnum 1.2 and 0.7 mm  data from Beuther et al. (2004a, 2005b). 
The squares are detections, the respective error bars were smaller than 
the squares and are not presented.  The triangles with arrows are upper limits
(4$\sigma$). The line is a least-squares power law fit (of the form $S_\nu\propto\nu^{\alpha}$) 
to each spectra. In the upper pannel, the 1.2 mm measurement is the peak emission of 
this source. }}
\label{fig2}
\end{center}
\end{figure}

%%%%%%%%%%%%%%%%%% SOURCE IRAS 18182-1433

\newpage
\begin{figure}
\vspace{-1.5cm}
\begin{center}
\scalebox{0.50}{\includegraphics{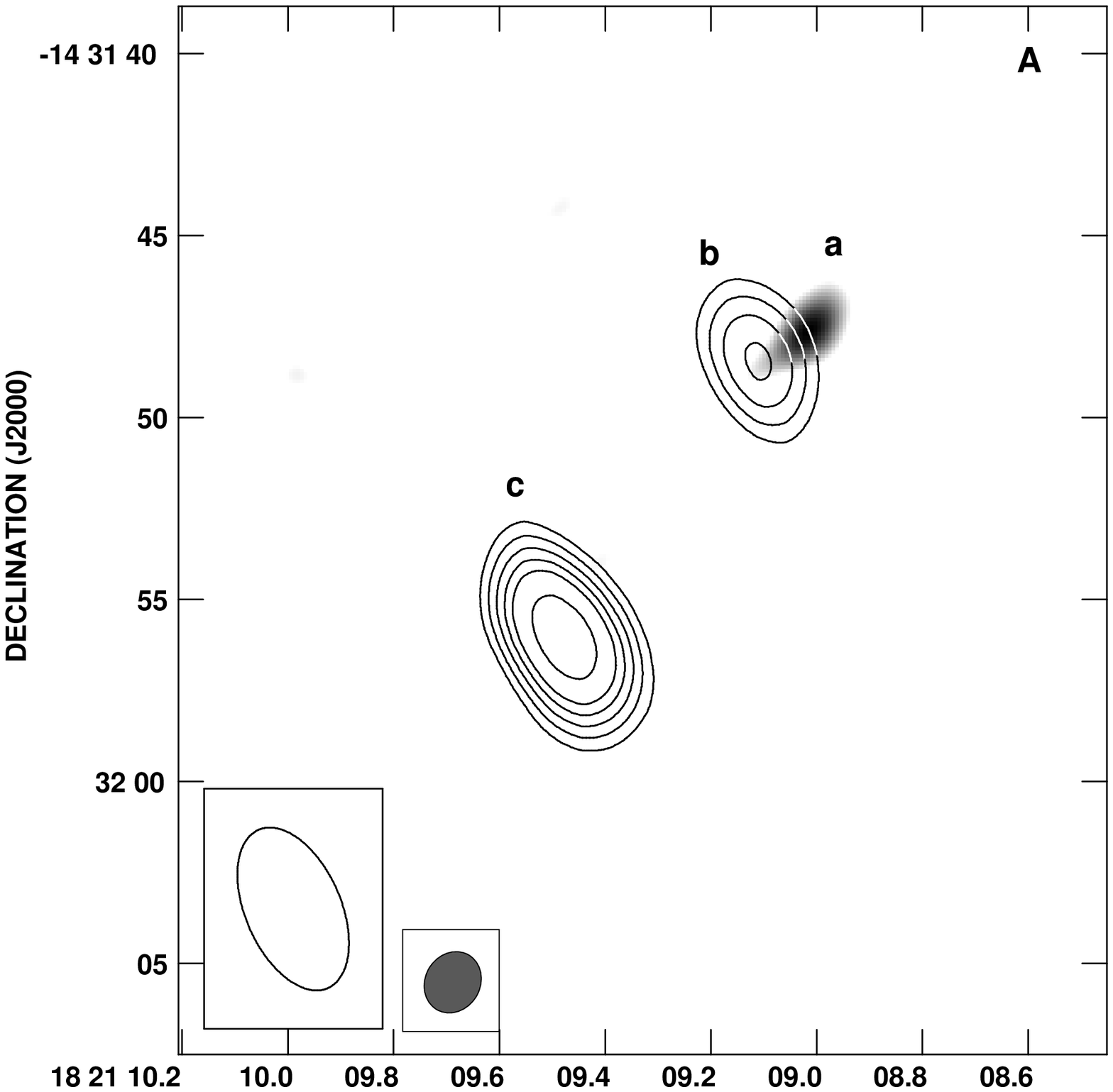}}\vspace{-1.7cm}\\ 
\scalebox{0.50}{\includegraphics{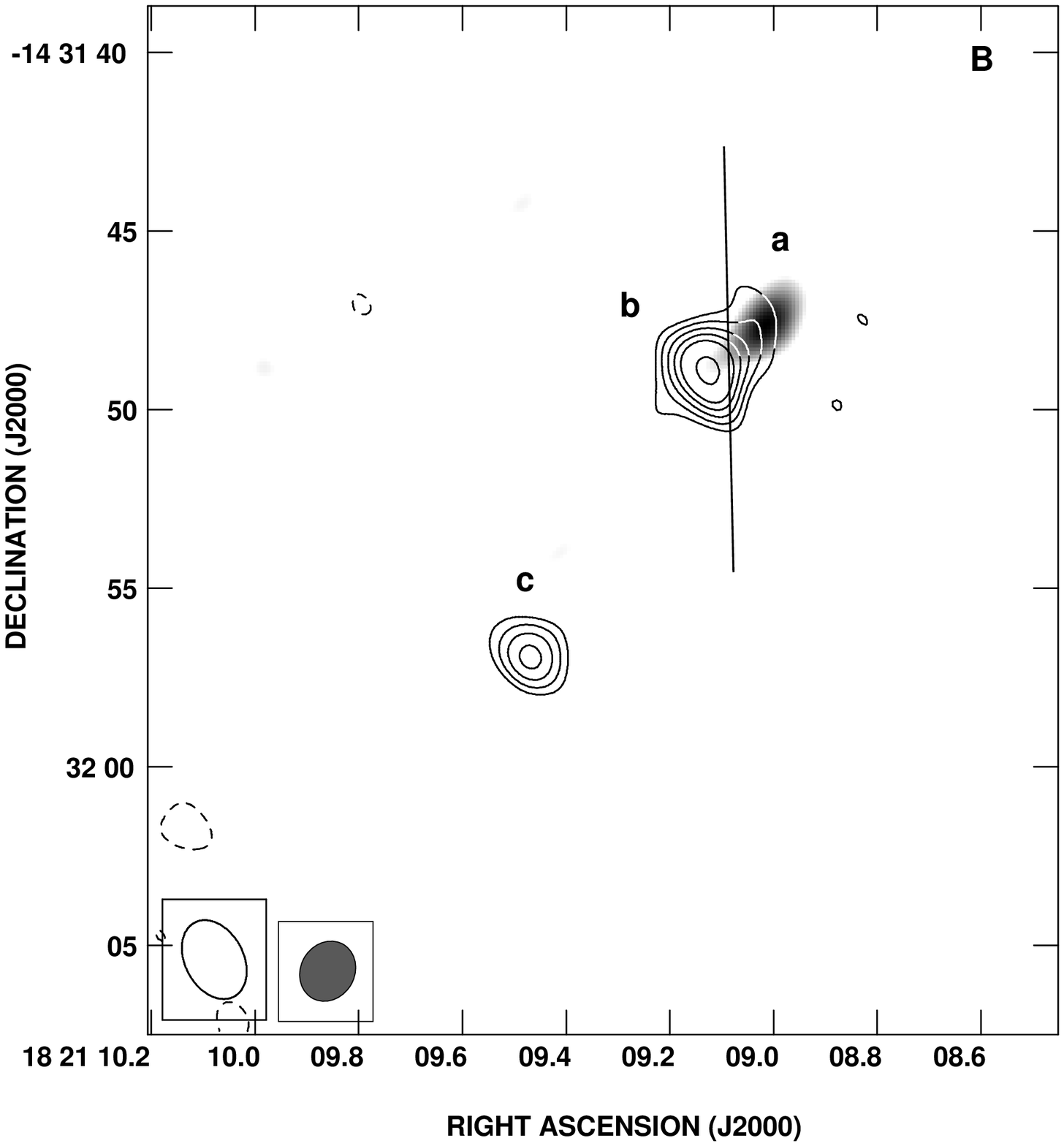}}\vspace{-0.9cm}\\
\vspace{-0.2cm}
\end{center}
\caption{{\scriptsize VLA continuum images of the source IRAS 18182-1433. 
{\bf A).} The contours show the 3.6 cm continuum emission and the grey scale shows the emission at 
7 mm.
The contours are -4, 4, 5, 6, 7, 8 and 10 times 55 $\mu$Jy beam$^{-1}$, the rms noise of the image.
{\bf B}) The contours show the 1.3 cm continuum emission and the grey scale shows the emission at 7 mm.
The contours are -4, 4, 5, 6, 7, 8 and 10 times 70 $\mu$Jy beam$^{-1}$, the rms noise of the image. 
The half power contour of the synthesized beam is shown in the bottom left corner of each image. 
The 7 mm continuum maps has a (u,v) tapering of 100 k$\lambda$. The line indicates the North-South  
CO(2-1) molecular outflow found by Beuther et al. (2002b). }}
\label{fig2.1}
\end{figure}
\newpage 

\begin{figure}
\vspace{-1.5cm}
\begin{center}
\scalebox{0.50}{\includegraphics{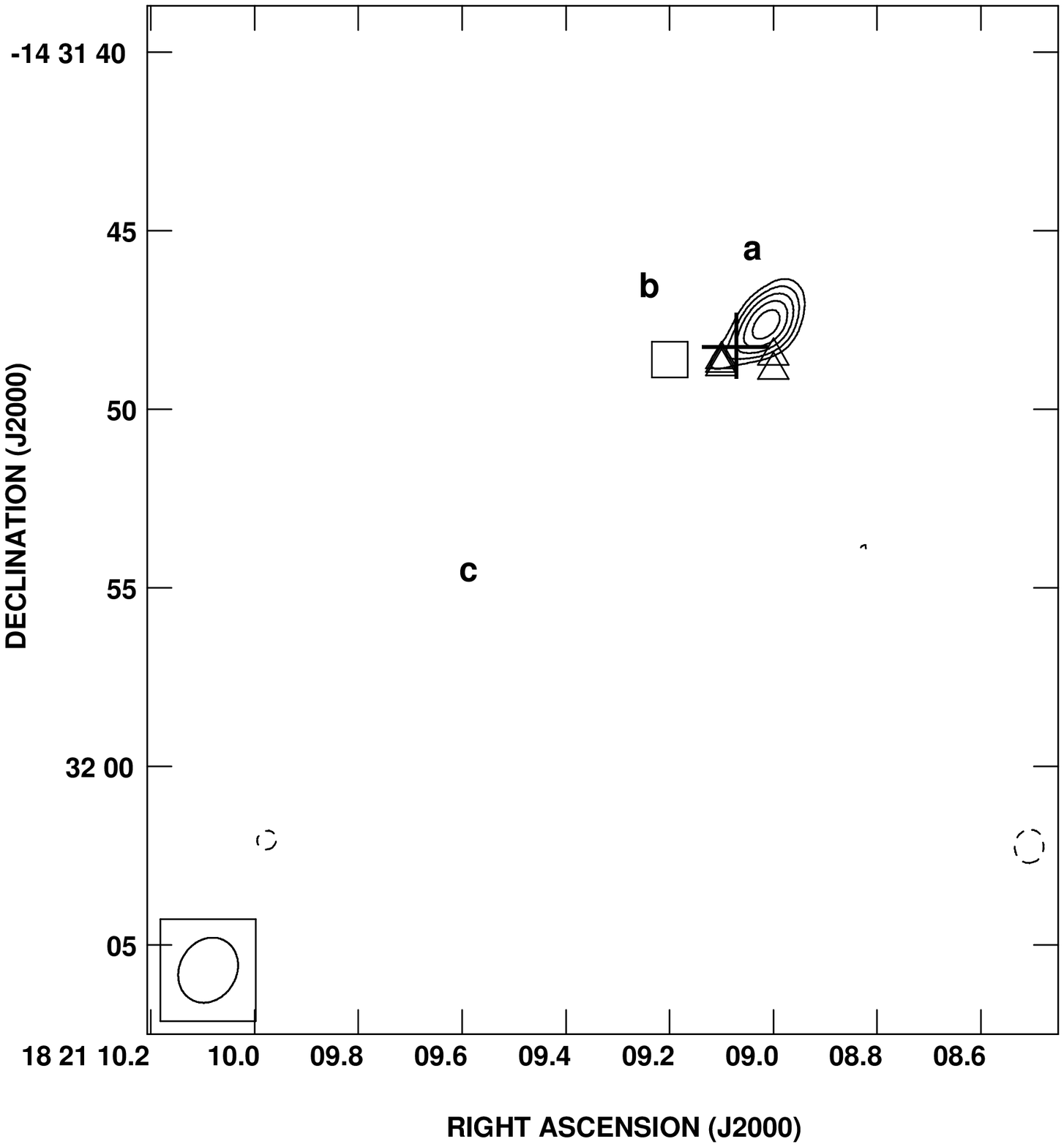}}\vspace{-0.5cm}\\
\end{center}
\caption{{\scriptsize VLA continuum image at 7 mm of the source IRAS 18182-1433. The half 
power contour of the synthesized beam is shown in the bottom left corner of the image.  
The contours are -4, 4, 5, 6, 7 and 8 times 0.22 mJy beam$^{-1}$, the rms noise of the image. 
The square and triangles indicate CH$_{3}$OH and H$_{2}$O maser positions 
(Walsh et al. 1998 and Beuther et al. 2002c), respectively. This 7 mm continuum map has a (u,v) 
tapering of 100 k$\lambda$. 
The cross indicates the 1.2 mm continnum emission peak found by Beuther et al. (2005), in preparation.}}
\label{fig2.2}
\end{figure}
\newpage 

\begin{figure}
\begin{center}
\scalebox{.96}{\includegraphics{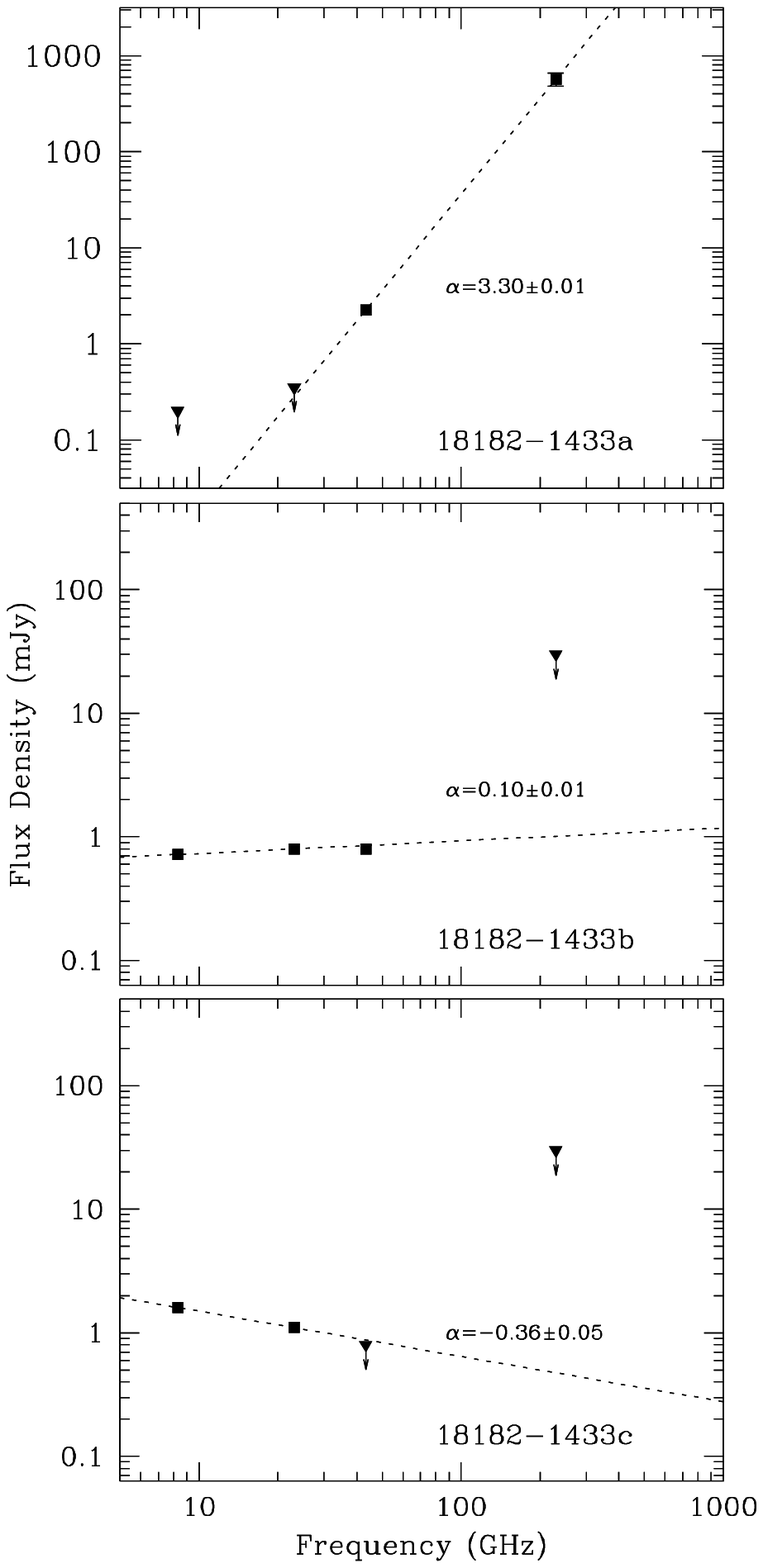}} 
\caption{{\scriptsize Spectral energy distribution (SED) for each detected radio continuum source in IRAS 
18182-1433 
combinig the 0.7, 1.3 and 3.6 cm VLA continuum data. The squares are detections, the respective error 
bars were smaller than the squares and are not presented. The triangles with arrows are upper  
limits (4$\sigma$). The line is a least-squares power law fit (of the form $S_\nu\propto\nu^{\alpha}$) to each 
spectra. }}
\label{fig4}
\end{center}
\end{figure}

%%%%%%%%%%%%%%%%%% SOURCE IRAS 18264-1152

\begin{figure}
\vspace{-1.5cm}
\begin{center}
\scalebox{.50}{\includegraphics{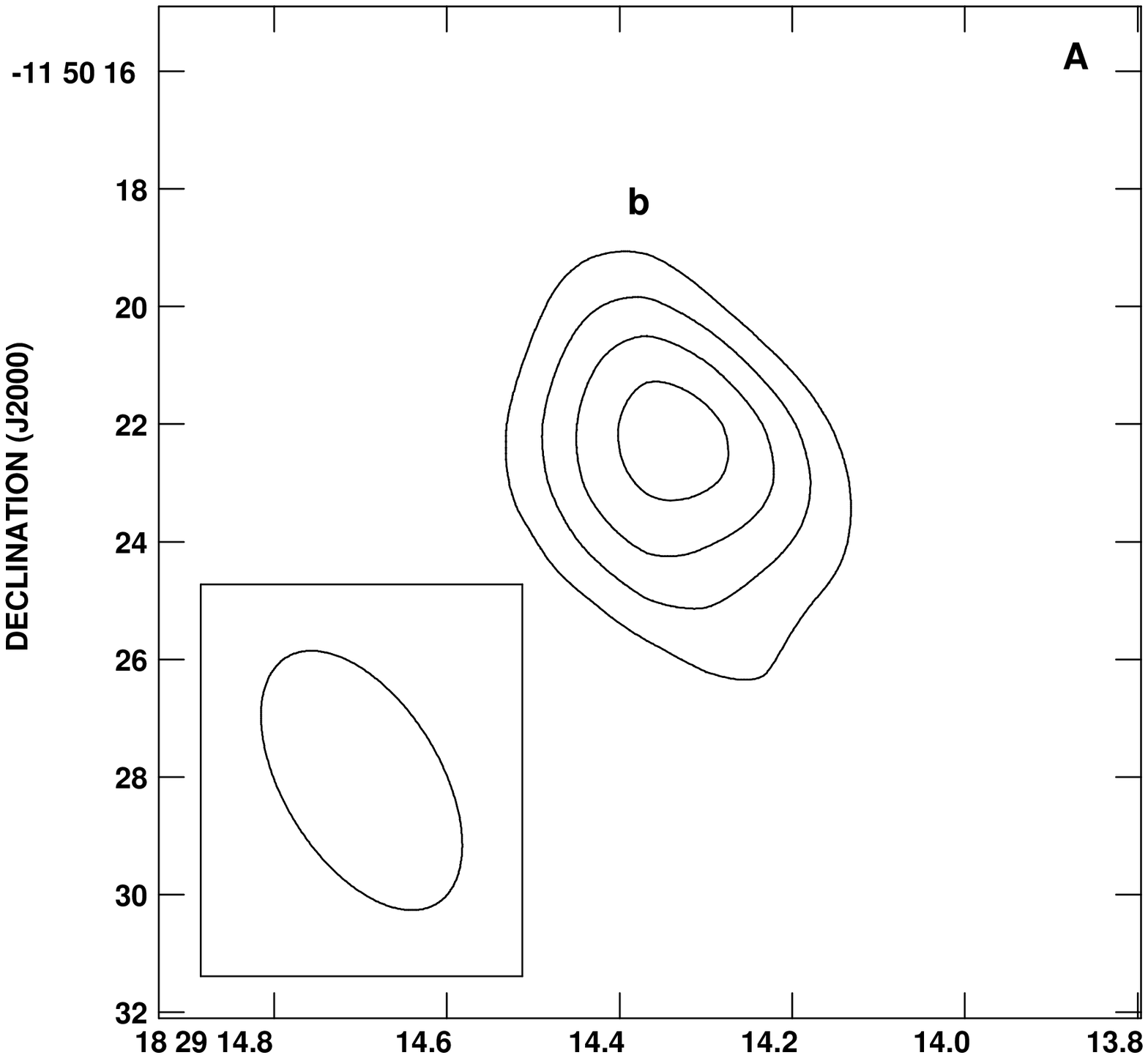}}\vspace{-1.7cm}\\ 
\scalebox{.50}{\includegraphics{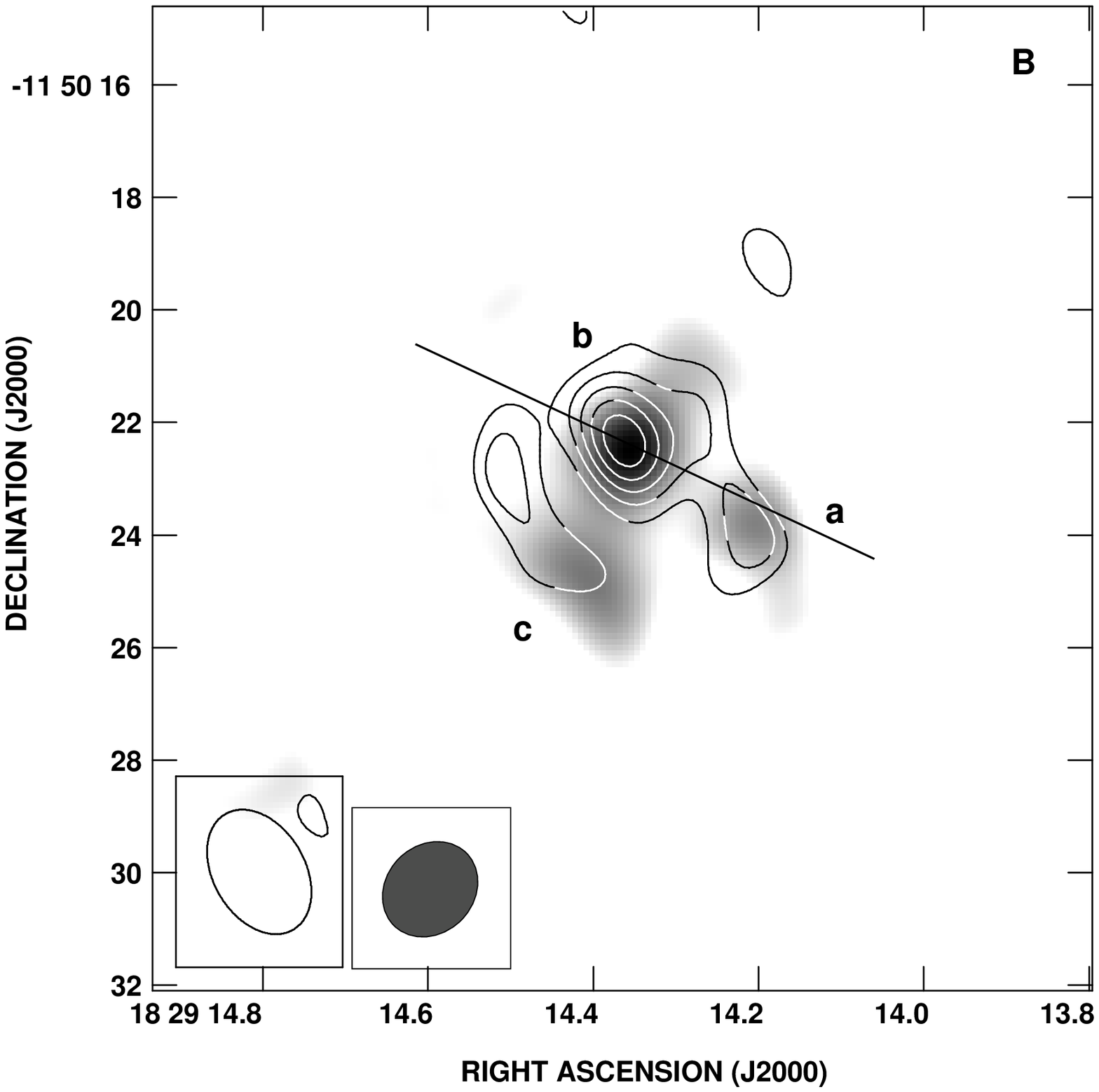}}\vspace{-1.0cm}\\
\end{center}
\caption{{\scriptsize VLA continuum images of the source IRAS 18264-1152.
{\bf A).} The 3.6 cm continuum emission. 
The contours are -4, 4, 6, 8 and 10 times 40 $\mu$Jy beam$^{-1}$, the rms noise of the image. 
{\bf B}) The contours show the 1.3 cm continuum emission and the grey scale shows the emission at 7 mm. 
The contours are -4, 4, 5, 6, 7 and 8 times 60 $\mu$Jy beam$^{-1}$, the rms noise of the image.
The half power contour of the synthesized beam is shown in the bottom left corner of each image. 
The 7 mm continuum map has a (u,v) tapering of 100 k$\lambda$. The line indicates the East-West 
CO(2-1) molecular outflow found by Beuther et al. (2002b).}}
\label{fig5}
\end{figure}
\newpage

\begin{figure}
\vspace{-0.5cm}
\begin{center}
\scalebox{0.50}{\includegraphics{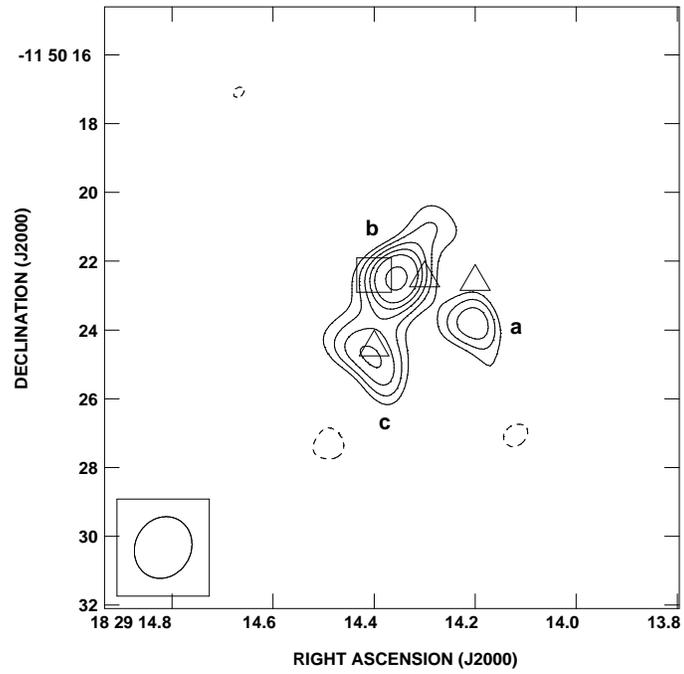}}\vspace{-0.5cm}\\
\end{center}
\caption{{\scriptsize VLA continuum image at 7 mm of the source IRAS 18264-1152. The half 
power contour of the synthesized beam is shown in the bottom left corner of the image. 
 The contours are -4, 4, 5, 6, 7, 8 and 10  times 0.18 mJy beam$^{-1}$, the rms noise of the image. 
 The square and triangles indicate CH$_{3}$OH and H$_{2}$O 
 masers positions (Beuther et al. 2002c), respectively. This 7 mm continuum map has a (u,v) tapering 
of 100 k$\lambda$. It is interesting to note that every 7 mm radio continuum source has 
associated a H$_{2}$O maser spot.
}}
\label{fig2.2}
\end{figure}

\newpage 
\begin{figure}
\begin{center}
\scalebox{.96}{\includegraphics{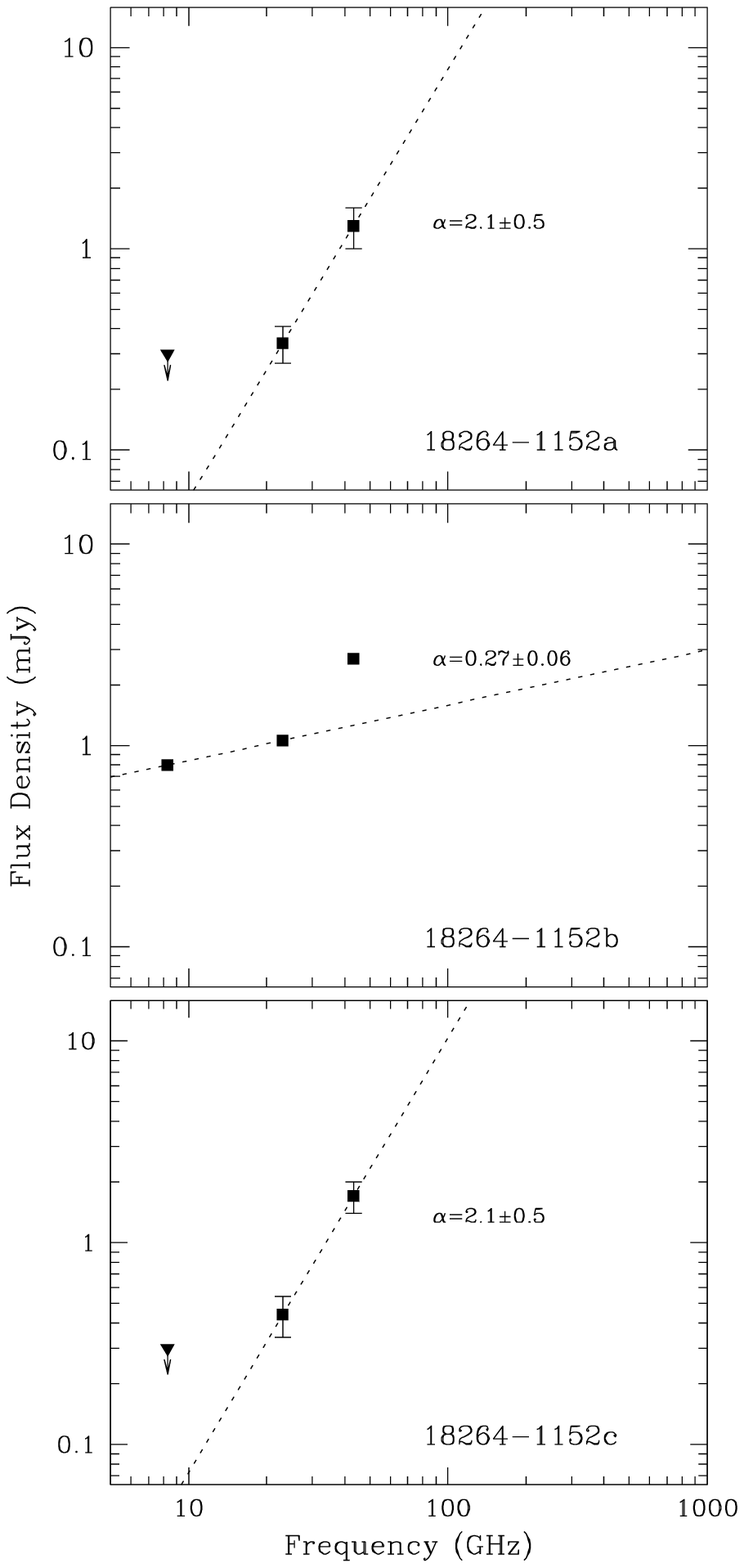}} 
\caption{\scriptsize Spectral energy distribution (SED) for each detected radio continuum source in IRAS 18264-1152
combining the 0.7, 1.3 and 3.6 cm VLA continuum data. The squares are detections, in a few cases the 
respective error bars were smaller than the squares and are not presented. The triangles with arrows 
are upper limits (4$\sigma$). The line is a least-squares power law fit
(of the form $S_\nu\propto\nu^{\alpha}$) to each spectra.}
\label{fig6}
\end{center}
\end{figure}

%%%%%%%%%%%%%%%%%% SOURCE IRAS 18308

\begin{figure} 
\begin{center}
\vspace{-0.5cm}
\scalebox{.50}{\includegraphics{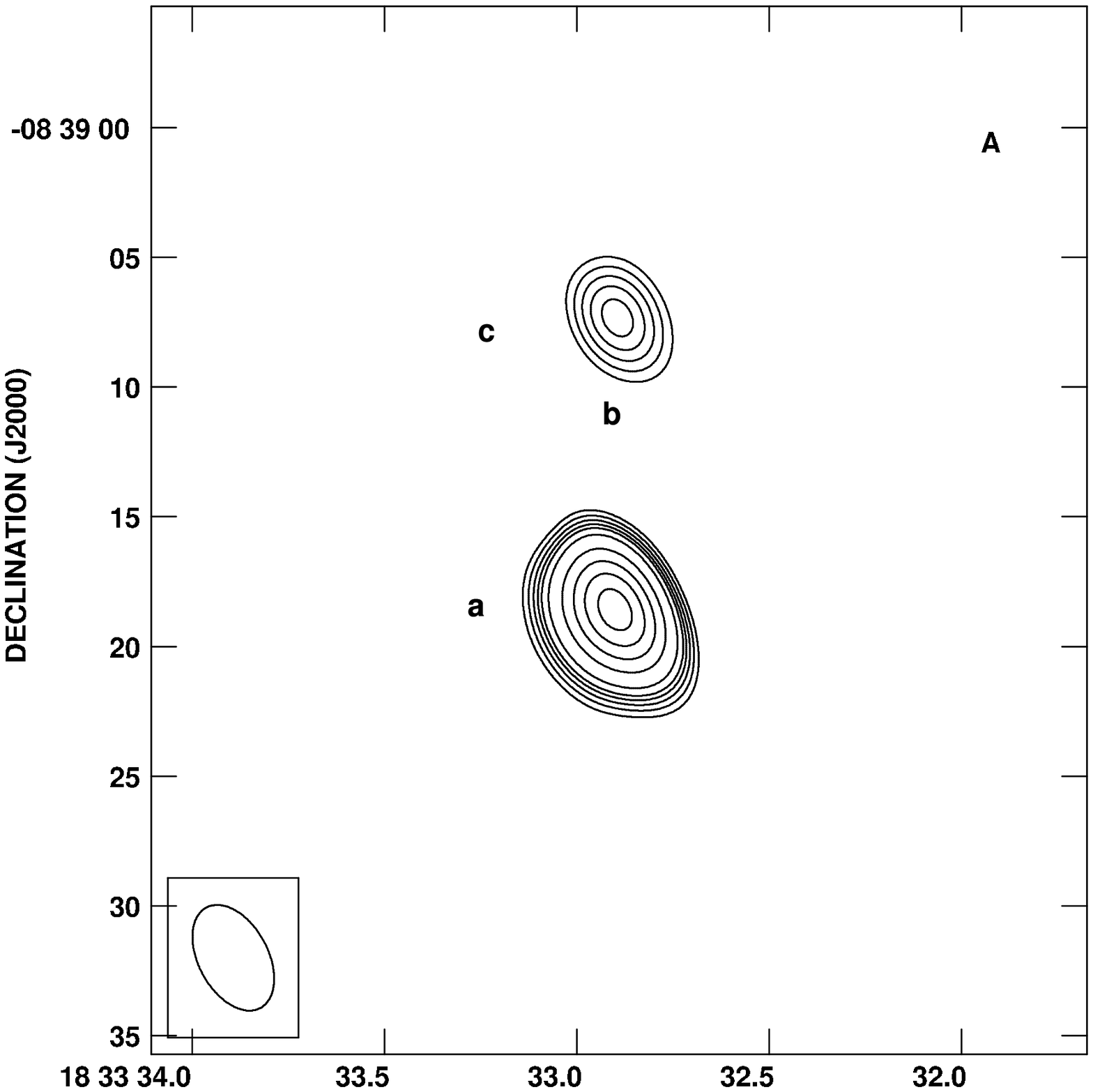}}
\vspace{-2.7cm}\\ 
\scalebox{.50}{\includegraphics{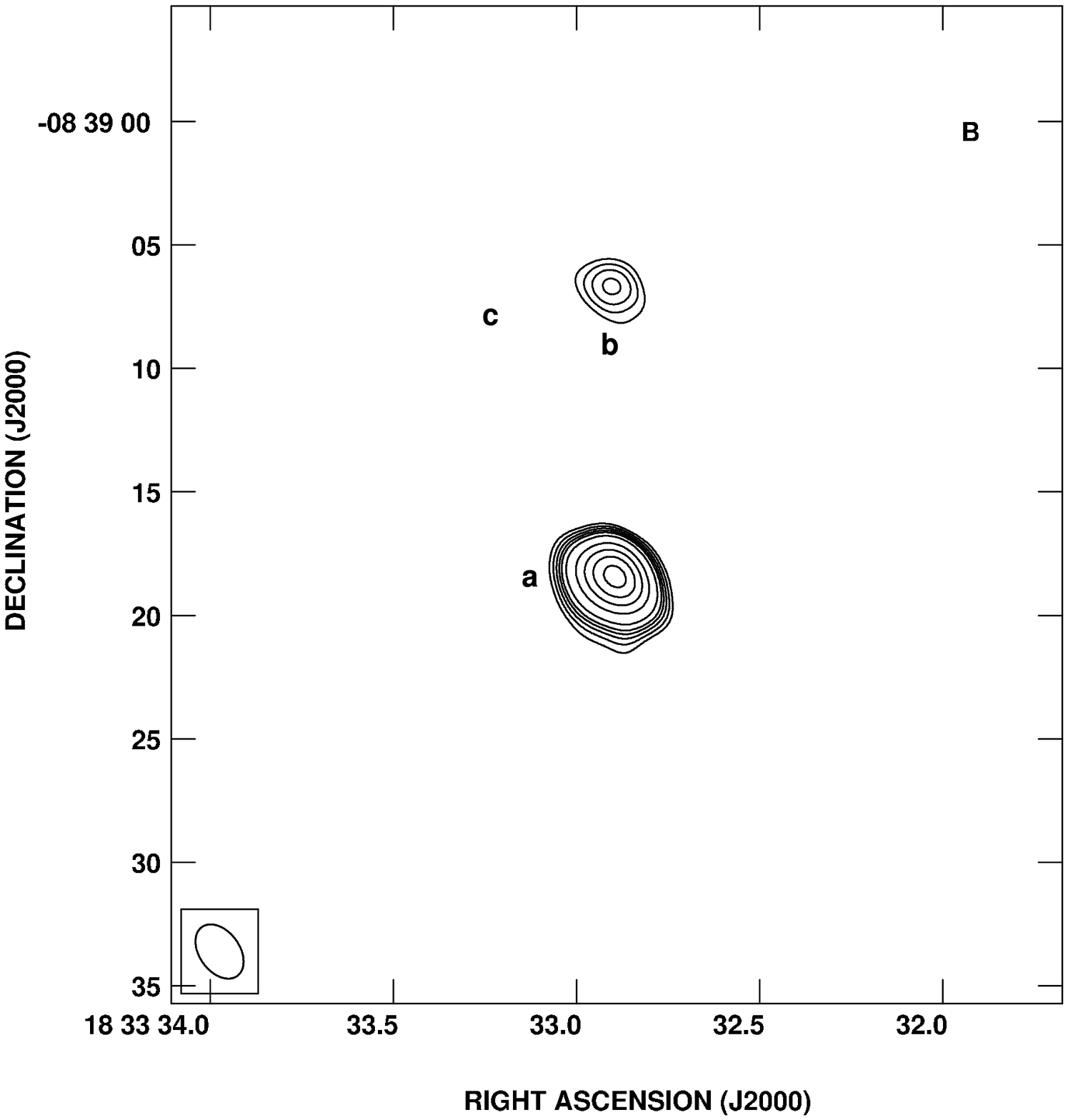}}
\vspace{-2.7cm}\\
\end{center}
\caption{{\scriptsize VLA continuum images at 3.6 ({\bf A}) and 1.3 ({\bf B}) cm of the source IRAS 18308-0841. The half 
power contour of the synthesized beam is shown in the bottom left corner of each image. {\bf A).} The 
contours are -4, 4, 6, 8, 10, 12, 14, 16, 18, 20 and 30 times 0.16 mJy beam$^{-1}$, 
the rms noise of the image. {\bf B).} The contours are -4, 4, 6, 8, 10, 12, 14, 16, 18, 20 and 30 times 
0.17 mJy beam$^{-1}$, the rms noise of the image.}}
\label{fig8}
\end{figure}
\newpage

\begin{figure}
\vspace{-0.5cm}
\begin{center}
\scalebox{0.50}{\includegraphics{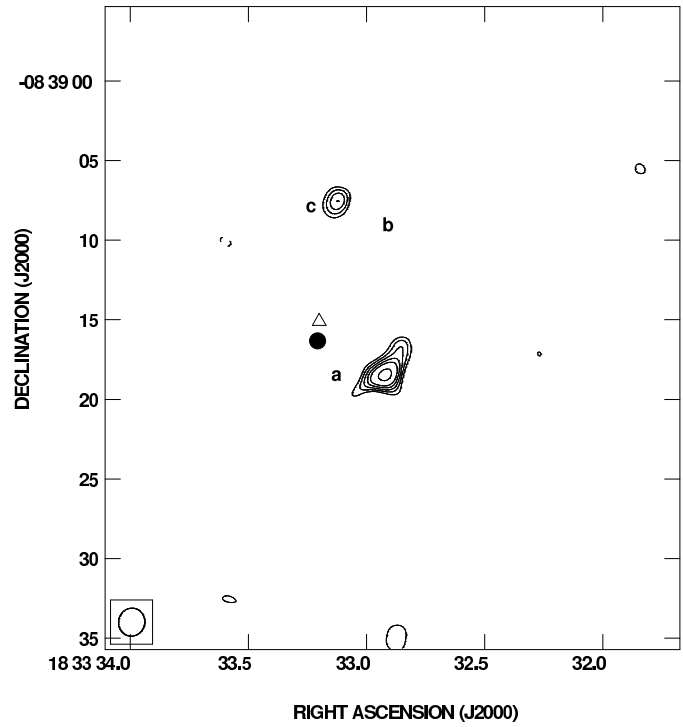}}\vspace{-2.5cm}\\
\end{center}
\caption{{\scriptsize VLA continuum image at 7 mm of the source IRAS 18308-0841. The half 
power contour of the synthesized beam is shown in the bottom left corner of the image.  
The contours are -4, 4, 5, 6, 7, 8 and 10 times 0.20 mJy beam$^{-1}$, the rms noise of the image. 
The triangle indicates the H$_{2}$O maser position (Beuther et al. 2002c).
This continuum map has a (u,v) tapering of 200 k$\lambda$. The dot indicates the MAMBO 1.2 mm continnum emission peak found by Beuther et al. (2002a)}}
\label{fig2.2}
\end{figure}
\newpage 

\begin{figure}
\begin{center}
\scalebox{.96}{\includegraphics{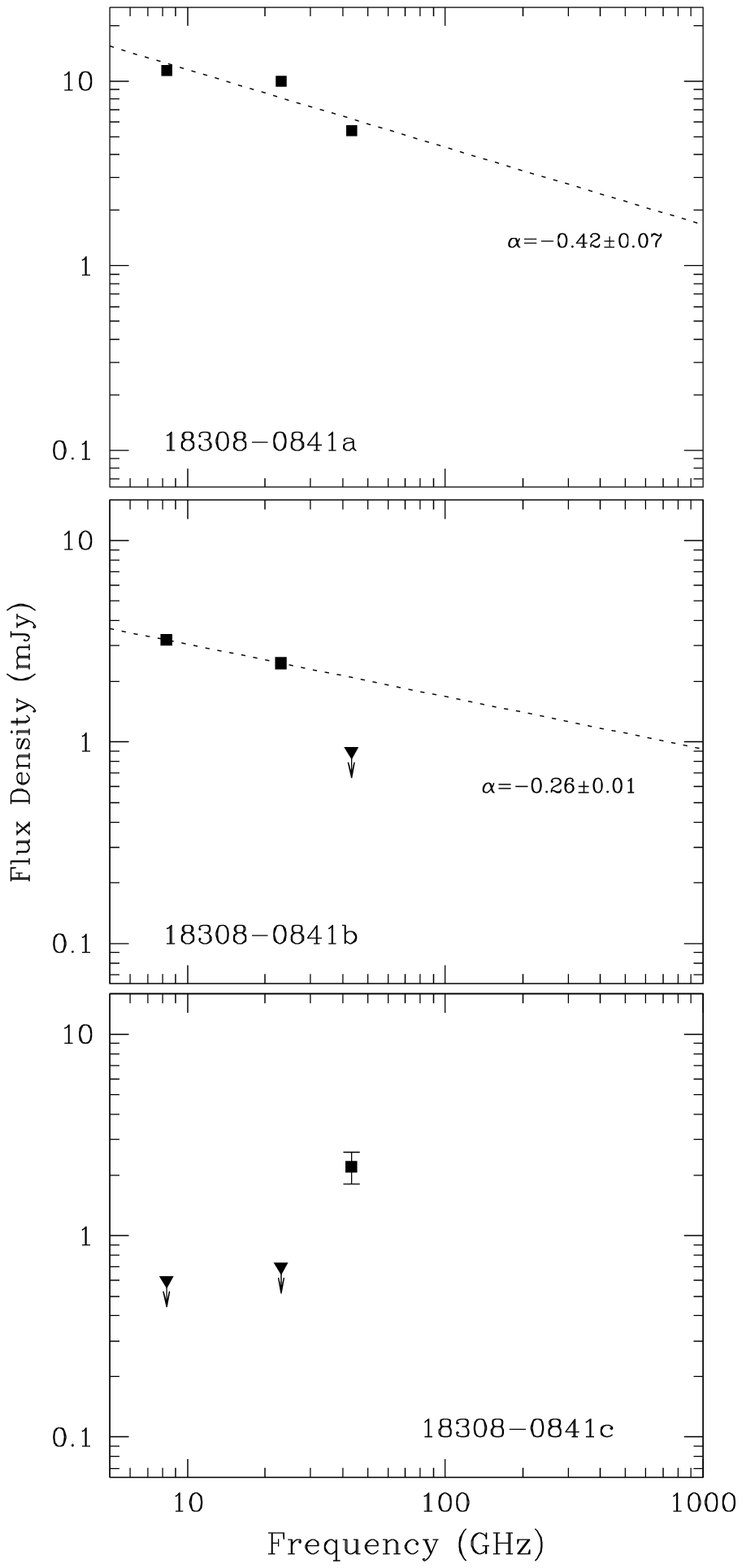}}
\caption{{\scriptsize Spectral energy distribution (SED) for each detected radio continuum source 
in IRAS 18308-0841 combining the 0.7, 1.3 and 3.6 cm VLA continuum data. 
The squares are detections, in 
most of the cases the respective error bars were smaller than the squares and 
are not presented. The triangles with arrows are upper limits (4$\sigma$).  The line is a 
least-squares power law fit (of the form $S_\nu\propto\nu^{\alpha}$) to each spectra.}}
\label{fig9}
\end{center}
\end{figure}

\newpage

%%%%%%%%%%%%%%%%%% COMETARY

\begin{figure}
\begin{center}
\scalebox{.86}{\includegraphics{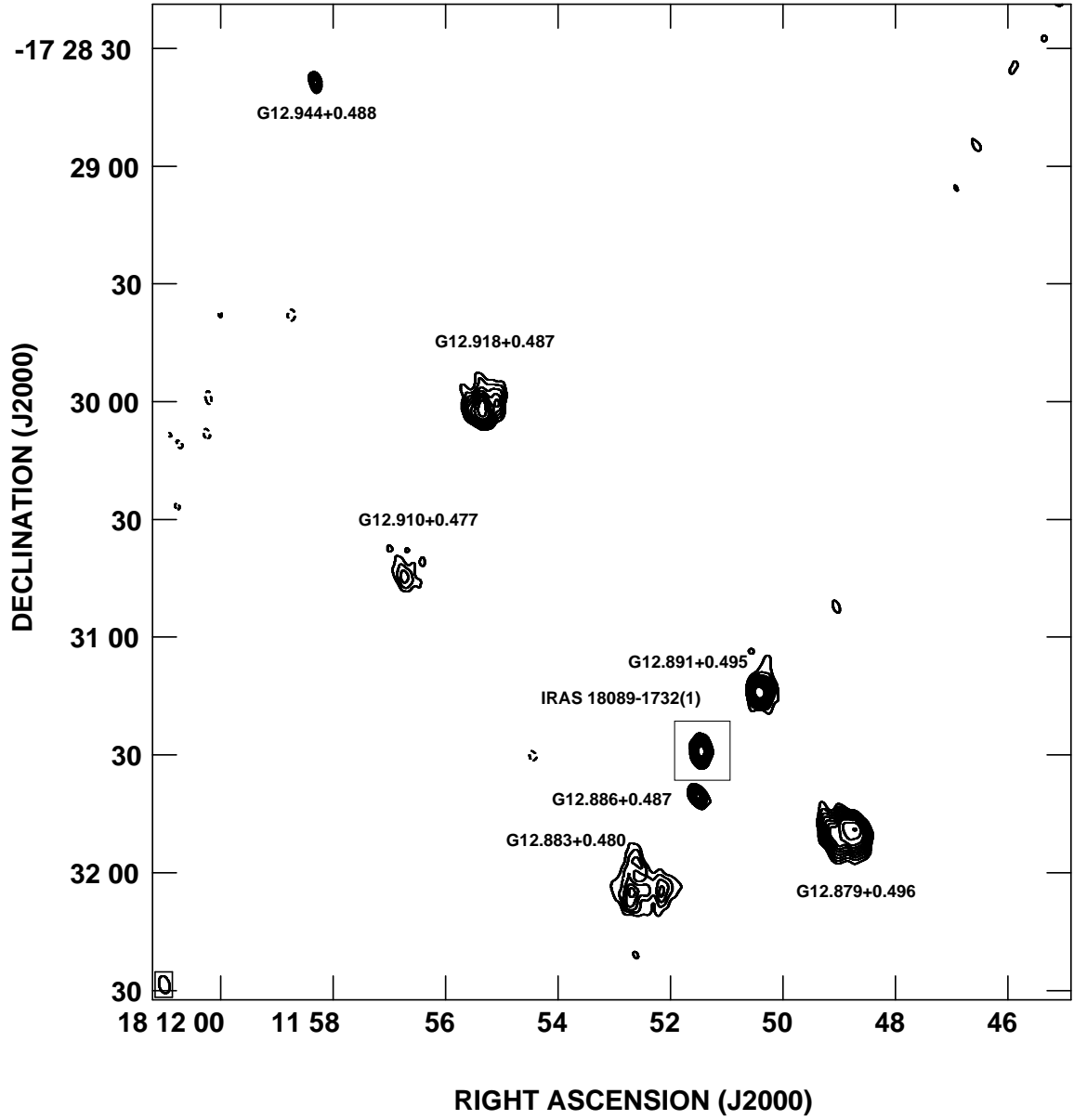}}
\caption{{\scriptsize VLA continuum image at 3.6 cm towards the source IRAS 18089-1732(1).
This map was made with ROBUST=5 (equivalent to natural weighting).
 The half power contour of the synthesized beam is shown in the bottom left corner of the image. 
The contours are -3, 3, 4, 5, 6, 7, 8, 9, 10, 12, 14, 16, 30 and 60 times 40 $\mu$Jy beam$^{-1}$, 
the rms noise of the image. The sources G12.879+0.496, G12.891+0.495, G12.886+0.487, 
G12.918+0.487, G12.918+0.487, G12.910+0.477, and G12.944+0.488 are first reported here.
This map was not corrected by primary beam response. The box indicates the approximate
size of the figure 1a. }}
\label{fig9}
\end{center}
\end{figure}

\begin{figure}
\begin{center}
\scalebox{.86}{\includegraphics{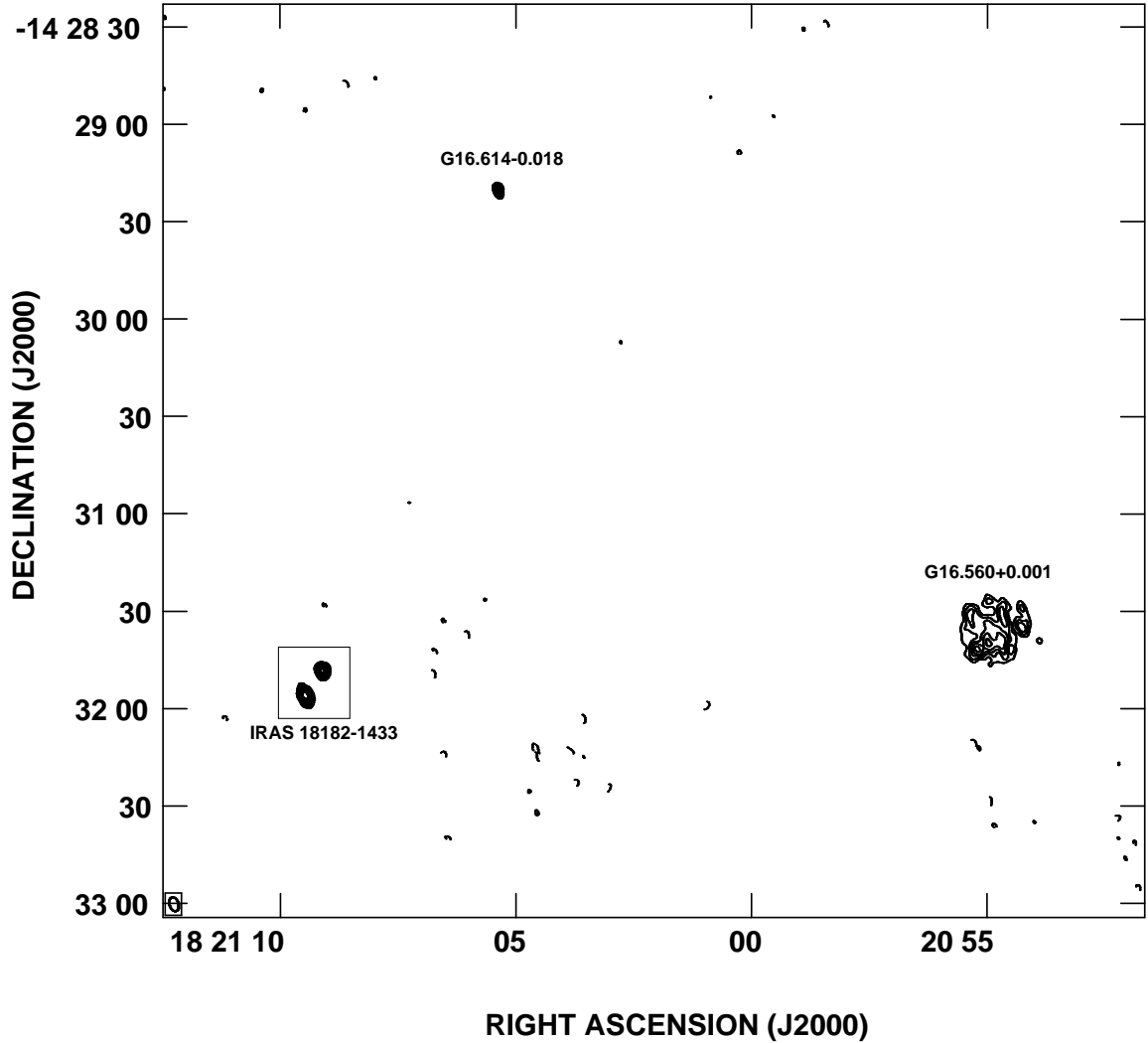}}
\caption{{\scriptsize VLA continuum image at 3.6 cm towards the source IRAS 181821433.
This map was made with ROBUST=5 (equivalent to natural weighting).
 The half power contour of the synthesized beam is shown in the bottom left corner of the image. 
The contours are -3, 3, 4, 5, 6, 7, 8, 9, 10, 12, 14, 16, 30 and 60 times 3.5 $\mu$Jy beam$^{-1}$, 
the rms noise of the image. The sources G16.560+0.001 and G16.614-0.018 are first reported here.
This map was not corrected by primary beam response. The box indicates the approximate
size of the figure 4a.}}
\label{fig9}
\end{center}
\end{figure}
\newpage

\begin{deluxetable}{l c c c c c c c}
\tablecolumns{9} 
\tablewidth{0pc} 
\tablecaption{Physical Parameters of the 10 Selected 1.2 mm Continuum Sources$^a$}
\tablehead{
\colhead{}                       &
\colhead{}                       &
\colhead{}                       &
\colhead{}                       &
\multicolumn{2}{c}{Physical Parameters} \\ 
\cline{4-6}
\colhead{}                       &
\multicolumn{2}{c}{Coordinates} &
\colhead{Distance$^b$}             &
\colhead{Luminosity} &
\colhead{Core Mass$^c$}        \\
\colhead{Source} & 
\colhead{$\alpha_{2000}$}  &
\colhead{$\delta_{2000}$}  &
\colhead{[Kpc]} &
\colhead{[\Lsun]}  &
\colhead{[\Msun]}  
} 
\startdata
IRAS 18089$-$1732(1)$d$& 18 11 51.3 & -17 31 29 &   3.6 & 3.2 $\times$ 10$^4$ & 1200 \\
IRAS 18089$-$1732(4)   & 18 12 30.4 & -17 32 59 &   3.6 &	       -     & 200  \\
IRAS 18090$-$1832      & 18 12 01.9 & -18 31 56 &   6.6 & 1.3 $\times$ 10$^4$  & 1000 \\
IRAS 18182$-$1433      & 18 21 07.9 & -14 31 53 &   4.5 & 2.0 $\times$ 10$^4$  & 1500 \\
IRAS 18264$-$1152      & 18 29 14.3 & -11 50 26 &   3.5 & 1.0 $\times$ 10$^4$  & 2200 \\
IRAS 18290$-$0924      & 18 31 44.8 & -09 22 09 &   5.3 & 2.5 $\times$ 10$^4$  & 800 \\
IRAS 18308$-$0841      & 18 33 31.9 & -08 39 17 &   4.9 & 1.6 $\times$ 10$^4$  & 1300 \\
IRAS 18521$+$0134      & 18 54 40.8 &  01 38 02 &   5.0 & 1.3 $\times$ 10$^4$  & 500  \\
IRAS 18553$+$0414      & 18 57 52.9 &  04 18 06 &   0.6 & 2.5 $\times$ 10$^2$  & 15   \\
IRAS 19012$+$0536      & 19 03 45.1 &  05 40 40 &   4.6 & 1.6 $\times$ 10$^4$  & 500 \\
\enddata
\tablecomments{\footnotesize Units of right ascension are hours, minutes, and seconds, 
                             and units of declination are degrees,
                             arcminutes, and arcseconds.}
\tablenotetext{a}{\footnotesize These parameters were obtained from Sridharan et al. (2002).}
\tablenotetext{b}{\footnotesize This distance listed is the nearest distance, and the luminosity
and core mass were calculated with this distance.}
\tablenotetext{c}{\footnotesize The masses listed here are a factor of two less than 
calculated by Beuther et al. (2002a). This difference is due to Beuther et al. (2002)a use 
a sightly different procedure to calculate the total mass of the cores that the developed by  
Hildebrand 1983. See the erratum for this paper.}
\tablenotetext{d}{\footnotesize
This source is previously known as IRAS 18089-1732, however we have added
a numbering (1) at the end of the name to distinguish between this
component and component IRAS 18089-1732(4)}

\end{deluxetable}

\pagebreak

\begin{deluxetable}{l c c c c c c c c}
\tablecolumns{9} 
\tablewidth{0pc} 
\tablecaption{Parameters of the VLA 0.7 cm Continuum Observations.}
\tablehead{
\colhead{}                       &
\colhead{}                       &
\colhead{}                       &
\colhead{}                       &
\colhead{Bootstrapped}           &
\colhead{Rms}                    &
\multicolumn{2}{c}{Synthesized Beam} \\ 
\cline{7-8}
\colhead{}                       &
\multicolumn{2}{c}{Phase Center} &
\colhead{Phase}             &
\colhead{Flux Density} &
\colhead{Noise}        &
\colhead{Size}         &
\colhead{P.A.}                \\
\cline{2-3} 
\colhead{Source} & 
\colhead{$\alpha_{2000}$}  &
\colhead{$\delta_{2000}$}  &
\colhead{Calibrator} &
\colhead{[Jy]}  &
\colhead{[mJy]}  &
\colhead{[arcsec]}   &
\colhead{[deg.] } 
} 
\startdata
18089$-$1732(1) & 18 11 51.3 & -17 31 29 & 1820-254 & 0.479$\pm$0.008  & 0.18 & 1.63$\times$1.08 & -29\\     
18089$-$1732(4) & 18 12 30.4 & -17 32 59 & 1820-254 & 0.479$\pm$0.008  & 0.16 & 2.11$\times$1.47 & -33\\
18090$-$1832    & 18 12 01.9 & -18 31 56 & 1820-254 & 0.479$\pm$0.008  & 0.18 & 1.99$\times$1.49 & -30\\     
18182$-$1433    & 18 21 07.9 & -14 31 53 & 1832-105 & 0.656$\pm$0.006  & 0.19 & 1.90$\times$1.59 & -30\\
18264$-$1152    & 18 29 14.3 & -11 50 26 & 1832-105 & 0.656$\pm$0.006  & 0.18 & 1.84$\times$1.62 & -28\\
18290$-$0924    & 18 31 44.8 & -09 22 09 & 1832-105 & 0.656$\pm$0.006  & 0.16 & 1.37$\times$1.14 & -10\\
18308$-$0841    & 18 33 31.9 & -08 39 17 & 1832-105 & 0.656$\pm$0.006  & 0.22 & 1.77$\times$1.64 & -14\\
18521$+$0134    & 18 54 40.8 &  01 38 02 & 1851+005 & 0.79$\pm$0.01    & 0.13 & 1.35$\times$1.18 & -9\\
18553$+$0414    & 18 57 52.9 &  04 18 06 & 1851+005 & 0.79$\pm$0.01    & 0.13 & 1.38$\times$1.20 & -22\\
19012$+$0536    & 19 03 45.1 &  05 40 40 & 1851+005 & 0.79$\pm$0.01    & 0.12 & 1.36$\times$1.19 & -30\\
\enddata
\tablecomments{\footnotesize Units of right ascension are hours, minutes, and seconds, 
                             and units of declination are degrees,
                             arcminutes, and arcseconds.}
\end{deluxetable}

\newpage

\begin{deluxetable}{l c c c c c c c c}
\tablecolumns{9} 
\tablewidth{0pc} 
\tablecaption{Parameters of the 1.3 cm VLA Continuum Observations.}	
\tablehead{
\colhead{}                       &
\colhead{}                       &
\colhead{}                       &
\colhead{}                       &
\colhead{Bootstrapped}           &
\colhead{Rms}           &
\multicolumn{2}{c}{Synthesized Beam} \\ 
\cline{7-8}
\colhead{}                       &
\multicolumn{2}{c}{Phase Center} &
\colhead{Phase}             &
\colhead{Flux Density} &
\colhead{Noise}        &
\colhead{Size}         &
\colhead{P.A.}                \\
\cline{2-3} 
\colhead{Source} & 
\colhead{$\alpha_{2000}$}  &
\colhead{$\delta_{2000}$}  &
\colhead{Calibrator} &
\colhead{[Jy]}  &
\colhead{[$\mu$Jy]}  &
\colhead{[arcsec]}   &
\colhead{[deg.] } 
}
\startdata
18089$-$1732(1)& 18 11 51.3 & -17 31 29 & 1820-254 & 0.626$\pm$0.003  & 70  & 1.73$\times$0.94 & 15\\     
18182$-$1433   & 18 21 07.9 & -14 31 53 & 1832-105 & 0.981$\pm$0.006  & 70  & 2.35$\times$1.63 & 28\\
18264$-$1152   & 18 29 14.3 & -11 50 26 & 1832-105 & 0.981$\pm$0.006  & 60  & 2.38$\times$1.63 & 31\\
18308$-$0841   & 18 33 31.9 & -08 39 17 & 1832-105 & 0.981$\pm$0.006  & 170 & 2.44$\times$1.63 & 36\\
\enddata
\tablecomments{\footnotesize Units of right ascension are hours, minutes, and seconds, 
                             and units of declination are degrees,
                             arcminutes, and arcseconds.}
\end{deluxetable}

\newpage

\begin{deluxetable}{l c c c c c c c}
\tablecolumns{9} 
\tablewidth{0pc} 
\tablecaption{Parameters of the VLA 3.6 cm Continuum Observations.}
\tablehead{
\colhead{}                       &
\colhead{}                       &
\colhead{}                       &
\colhead{}                       &
\colhead{Bootstrapped}           &
\colhead{Rms}           &
\multicolumn{2}{c}{Synthesized Beam} \\ 
\cline{7-8}
\colhead{}                       &
\multicolumn{2}{c}{Phase Center} &
\colhead{Phase}             &
\colhead{Flux Density} &
\colhead{Noise}        &
\colhead{Size}         &
\colhead{P.A.}                \\
\cline{2-3} 
\colhead{Source} & 
\colhead{$\alpha_{2000}$}  &
\colhead{$\delta_{2000}$}  &
\colhead{Calibrator} &
\colhead{[Jy]}  &
\colhead{[$\mu$Jy]}  &
\colhead{[arcsec]}   &
\colhead{[deg.] } 
}
\startdata
18089$-$1732(1)& 18 11 51.3 & -17 31 29 & 1820-254 & 0.771$\pm$0.002  & 40  & 4.70$\times$2.75 & 17\\     
18182$-$1433   & 18 21 07.9 & -14 31 53 & 1832-105 & 1.415$\pm$0.004  & 55  & 4.75$\times$2.62 & 24\\
18264$-$1152   & 18 29 14.3 & -11 50 26 & 1832-105 & 1.415$\pm$0.004  & 40  & 4.90$\times$2.66 & 31\\
18308$-$0841   & 18 33 31.9 & -08 39 17 & 1832-105 & 1.415$\pm$0.004  & 160 & 4.40$\times$2.68 & 28\\
\enddata
\tablecomments{\footnotesize Units of right ascension are hours, minutes, and seconds, 
                             and units of declination are degrees,
                             arcminutes, and arcseconds.}
\end{deluxetable}

\newpage

%\begin{sideways}
\begin{center}
\begin{deluxetable}{l c c c c c c}
\tablewidth{0pt}
\tabletypesize{\scriptsize}
%\small
\tablecaption{Physical Parameters of the HII Regions Detected with the VLA at 3.6 cm$^a$.}
\tablehead{
\colhead{}                       &
\colhead{}                       &
\colhead{}                       &
\multicolumn{4}{c}{Physical Parameters} \\
\cline{4-7}                  
\colhead{}                       &
\colhead{}                       &           
\colhead{}                       &                            
\colhead{}                       &
\colhead{}                       &
\colhead{Flux Density}           &
\colhead{log(N$_{L}$)$^b$} \\              
\colhead{Source}                 &
\colhead{$\alpha_{2000}$}        &
\colhead{$\delta_{2000}$}        &
\colhead{Deconvolved Size}       &
\colhead{P.A.}                   &
\colhead{[mJy]}                  &
\colhead{s$^{-1}$}                                                       
}
\startdata            
G12.879+0.496 & 18 11 48.834 & -17 31 49.07 &  $7\rlap.{''}4$ $\pm$ $0\rlap.{''}5$ $\times$ $5\rlap.{''}2$ $\pm$ $0\rlap.{''}5$ & 75$^{\circ}$ $\pm$ 10$^{\circ}$ & 6.0$\pm$0.1 &45.7\\
G12.891+0.495 & 18 11 50.422 & -17 31 14.12 &  $3\rlap.{''}0$ $\pm$ $0\rlap.{''}3$ $\times$ $2\rlap.{''}3$ $\pm$ $0\rlap.{''}6$ & 115$^{\circ}$ $\pm$68$^{\circ}$& 3.0$\pm$0.1 & 45.5\\
G12.886+0.487 & 18 11 51.510 & -17 31 40.57 &  $6\rlap.{''}4$ $\pm$ $1\rlap.{''}4$ $\times$ $3\rlap.{''}7$ $\pm$ $1\rlap.{''}5$ & 58$^{\circ}$ $\pm$ 25$^{\circ}$ & 1.0$\pm$0.1 &45.0\\
G12.883+0.480 & 18 11 52.680 & -17 32 04.92 &  $9\rlap.{''}0$ $\pm$ $0\rlap.{''}8$ $\times$ $8\rlap.{''}0$ $\pm$ $0\rlap.{''}7$ & 16$^{\circ}$ $\pm$ 50$^{\circ}$ & 6.0$\pm$0.1 &45.7\\
G12.918+0.487 & 18 11 55.360 & -17 30 01.92 &  $6\rlap.{''}0$ $\pm$ $0\rlap.{''}5$ $\times$ $5\rlap.{''}3$ $\pm$ $0\rlap.{''}4$ & 44$^{\circ}$ $\pm$ 26$^{\circ}$ & 6.0$\pm$0.1 &45.7\\
G12.910+0.477 & 18 11 56.751 & -17 30 44.05 &  $9\rlap.{''}6$ $\pm$ $1\rlap.{''}1$ $\times$ $7\rlap.{''}0$ $\pm$ $0\rlap.{''}9$ & 42$^{\circ}$ $\pm$ 20$^{\circ}$ & 5.0$\pm$0.1 &45.6\\
G12.944+0.488 & 18 11 58.314 & -17 28 38.78 &  $3\rlap.{''}6$ $\pm$ $0\rlap.{''}6$ $\times$ $1\rlap.{''}6$ $\pm$ $0\rlap.{''}7$ & 43$^{\circ}$ $\pm$ 13$^{\circ}$ & 1.0$\pm$0.1 &45.1\\
G16.560+0.001 & 18 20 54.977 & -14 31 39.84 &  -  &  - & 2.5$\pm$0.1 & 45.5 \\ 
G16.614-0.018 & 18 21 05.386 & -14 29 20.45  &  -  &  - & 0.9$\pm$0.1 & 45.1  \\
\enddata
\tablecomments{\footnotesize Units of right ascension are hours, minutes, and seconds, and units of declination are degrees, arcminutes, and arcseconds.}
\tablenotetext{a}{\footnotesize These parameters were obtained from a map with the primary beam response correction applied.}
\tablenotetext{b}{\footnotesize The flux of Lyman continuum photons N$_{L}$ was obtained assuming distances of 3.6 kpc and 4.5 kpc.}
\end{deluxetable}
\end{center}

%\begin{sideways}
\begin{deluxetable}{l c c c c c c c}
\tablewidth{0pt}
\tabletypesize{\scriptsize}
%\small
\tablecaption{Flux Densities at 0.7, 1.3 and 3.6 cm of the VLA Continuum Sources.}
\tablehead{
\colhead{}                       &
\colhead{}                       &
\colhead{}                       &
\multicolumn{3}{c}{Flux Density} &
\colhead{}\\
\cline{4-6}                      
\colhead{}                       &
\colhead{}                       &
\colhead{}                       &           
\colhead{3.6 cm}                 &                            
\colhead{1.3 cm}                 &
\colhead{0.7 cm}                 &
\colhead{} \\               
\colhead{Source}                 &
\colhead{$\alpha_{2000}$}        &
\colhead{$\delta_{2000}$}        &
\colhead{[mJy]}                  &
\colhead{[mJy]}                  &
\colhead{[mJy]}                  &                       
\colhead{Nature}                 
}
\startdata            
18089$-$1732(1)a & 18 11 51.451 & -17 31 28.85 &  1.10$\pm$0.05    &  1.60$\pm$0.04  &  3.00$\pm$0.30 & Thermal Jet + Core \\
18089$-$1732(1)b & 18 11 51.571 & -17 31 27.89 &    $\leq$ 0.20    &  $\leq$ 0.30   &  2.10$\pm$0.30 & H II Region?\\
18182$-$1433a & 18 21 09.014 & -14 31 47.62 &    $\leq$ 0.20    &  $\leq$ 0.35     &  2.26$\pm$0.30 & Core+Disk\\
18182$-$1433b & 18 21 09.129 & -14 31 48.59 &  0.60$\pm$0.05    &  0.80$\pm$0.05   &    0.72$\pm$0.10  &  Thermal Jet \\
18182$-$1433c & 18 21 09.469 & -14 31 56.93 &  1.65$\pm$0.03    &  1.10$\pm$0.05   &   $\leq$0.80  &  Strong Shocks? \\
18264$-$1152a & 18 29 14.203 & -11 50 23.80 &    $\leq$ 0.20    &  0.34$\pm$0.10   &  1.30$\pm$0.20  & H II Region?\\
18264$-$1152b & 18 29 14.356 & -11 50 22.50 &  0.80$\pm$0.02    &  1.06$\pm$0.07   &  2.70$\pm$0.20 &  Thermal Jet + Core\\
18264$-$1152c & 18 29 14.406 & -11 50 24.77 &    $\leq$ 0.20    &  0.44$\pm$0.10   &  1.70$\pm$0.30  & H II Region?\\
18308$-$0841a & 18 33 32.891 & -08 39 18.47 & 11.40$\pm$0.05    & 10.30$\pm$0.20   &  5.38$\pm$0.50  & Strong Shocks?\\
18308$-$0841b & 18 33 32.904 & -08 39 07.33 &  3.21$\pm$0.05    &  2.46$\pm$0.10   &    $\leq$0.80  &  Strong Shocks?\\
18308$-$0841c & 18 33 33.120 & -08 39 07.53 &    $\leq$ 0.60    &  $\leq$ 0.75     &  2.20$\pm$0.40  & Core?\\
\enddata
\tablecomments{\footnotesize Units of right ascension are hours, minutes, and seconds, and 
units of declination are degrees, arcminutes, and arcseconds.}
\end{deluxetable}
%\end{sideways}
\newpage

\end{document}